\documentclass[aps,prl,10pt,floatfix,superscriptaddress,noeprint,reprint]{revtex4-2}
\usepackage{amsmath,amsfonts,amssymb,physics,graphicx}
\usepackage[dvipsnames]{xcolor}
\usepackage[colorlinks=true,
hyperfootnotes=true,
bookmarks=true,
breaklinks=true,
citecolor=blue,
urlcolor=blue,
linkcolor=blue,
]{hyperref}
\newcommand{\como}{Center for Nonlinear and Complex Systems, Dipartimento di Scienza e Alta Tecnologia, Università degli Studi dell'Insubria, Via Valleggio 11, 22100 Como, Italy}
\newcommand{\infnM}{INFN, Sezione di Milano, 20133 Milano, Italy}
\newcommand{\infnP}{INFN, Sezione di Pavia, 27100 Pavia, Italy}
\newcommand{\infnC}{INFN, Sezione di Catania, 95123 Catania, Italy}
\newcommand{\catania}{Dipartimento di Fisica e Astronomia ‘Ettore Majorana’, Università di Catania, Via S. Sofia 64, 95123 Catania, Italy}
\newcommand{\pavia}{Dipartimento di Fisica ‘Alessandro Volta’, Università di Pavia, Via Bassi 6, 27100 Pavia, Italy}

\begin{document}
\title{Engineering Nonclassical States via the Dynamical Casimir Effect}
\author{Maristella Crotti}
\affiliation{\como}
\affiliation{\infnM}
\author{Luca Razzoli}
\affiliation{\pavia}
\affiliation{\infnP}
\author{Giacomo Guarnieri}
\affiliation{\pavia}
\affiliation{\infnP}
\author{Luigi Giannelli}
\affiliation{\catania}
\affiliation{\infnC}
\author{Giuseppe A. Falci}
\affiliation{\catania}
\affiliation{\infnC}
\author{Giuliano Benenti}
\affiliation{\como}
\affiliation{\infnM}

\date{\today}

\begin{abstract}
Nonadiabatic driving in ultrastrongly coupled light--matter systems is commonly regarded as a source of errors, as counter-rotating interactions convert vacuum fluctuations into real excitations through the dynamical Casimir effect (DCE). Here we show that, instead, the DCE can be harnessed as a resource for engineering nonclassical states of light. Considering a cavity mode ultrastrongly coupled to a frequency-tunable qubit, we employ optimal quantum control to design driving protocols that convert vacuum fluctuations into targeted states. Numerical optimization reveals a versatile and robust approach for the deterministic preparation of a broad class of nonclassical states, illustrated here through Fock states, squeezed states, and Schr\"{o}dinger-cat-state superpositions. 
\end{abstract}

\maketitle

\textit{Introduction.}---The generation of nonclassical states of light is a key objective in quantum technologies, with applications in quantum information processing and quantum-enhanced sensing \cite{Nielsen_Chuang_2010,BenentiCasati}. Squeezed states enable measurements beyond the standard quantum limit \cite{Walls2008, chen2021prl}, as exemplified by gravitational-wave detectors \cite{Aasi2013}, whereas Fock states constitute a fundamental resource for bosonic quantum computing~\cite{Descamps26}. Recent studies have shown that bosonic modes initialized in Fock states can even provide the energy required to implement arbitrary quantum gates through energy-recycling mechanisms \cite{kurman2026prx}. More generally, superpositions of Fock states and Schr\"{o}dinger-cat states form the basis of several bosonic error-correction and logical-encoding schemes \cite{michael2016prx,chen2021prr,PhysRevLett.111.120501,Ofek2016nature}.

The ability to rapidly prepare nonclassical states before decoherence degrades their quantum properties is a key requirement. Circuit quantum electrodynamics (QED) provides a promising platform for this purpose, allowing access to the ultrastrong light--matter coupling regime (USC) \cite{Beaudoin2011,Kockum2019,forndiaz2019}, where the interaction strength becomes comparable to the natural frequencies of the uncoupled subsystems. The resulting fast light--matter dynamics open new opportunities for the efficient engineering of nonclassical states.

However, pushing state-generation protocols into the USC regime inevitably brings the dynamical Casimir effect (DCE) into play~\cite{Moore1970JMathPhys,Dodonov_2010,Wilson2011,RevModPhys.84.1,Hoeb2017},  since nonadiabatic modulation of system parameters generates pairs of excitations from the vacuum. To date, the DCE has been predominantly regarded as an unwanted effect that limits the fidelity of quantum operations~\cite{PhysRevA.90.052313}. A notable exception is the pioneering proposal to exploit it as a resource for generating entangled states of superconducting qubits~\cite{felicetti2014prl}.

Here, we exploit the DCE as a resource for the fast and deterministic generation of nonclassical cavity states. To this end, we consider a single-mode electromagnetic cavity coupled to a two-level system (qubit) in the USC regime. In this regime, counter-rotating terms play a crucial role, enabling photon generation from the vacuum through nonadiabatic modulation of the qubit frequency, a manifestation of the parametric DCE 
\cite{Dodonov_2010}.
By employing optimal-control techniques \cite{Koch2022,PhysRevA.84.022326,Muller2022,KHANEJA2005296}, we design driving protocols that steer the system toward desired nonclassical states by maximizing the final-state fidelity. We illustrate the versatility of the approach through the generation of representative target states, including Fock, squeezed, and Schr\"{o}dinger-cat-state superpositions. We systematically assess its performance in terms of fidelity and control cost over a broad range of state parameters. Furthermore, we demonstrate its robustness against both classical control noise and quantum dissipation induced by a thermal environment. Our results establish a viable framework for harnessing the DCE as a resource for the fast and robust generation of high-fidelity nonclassical cavity states.

\begin{figure}
    \centering
    \includegraphics[width=\linewidth]{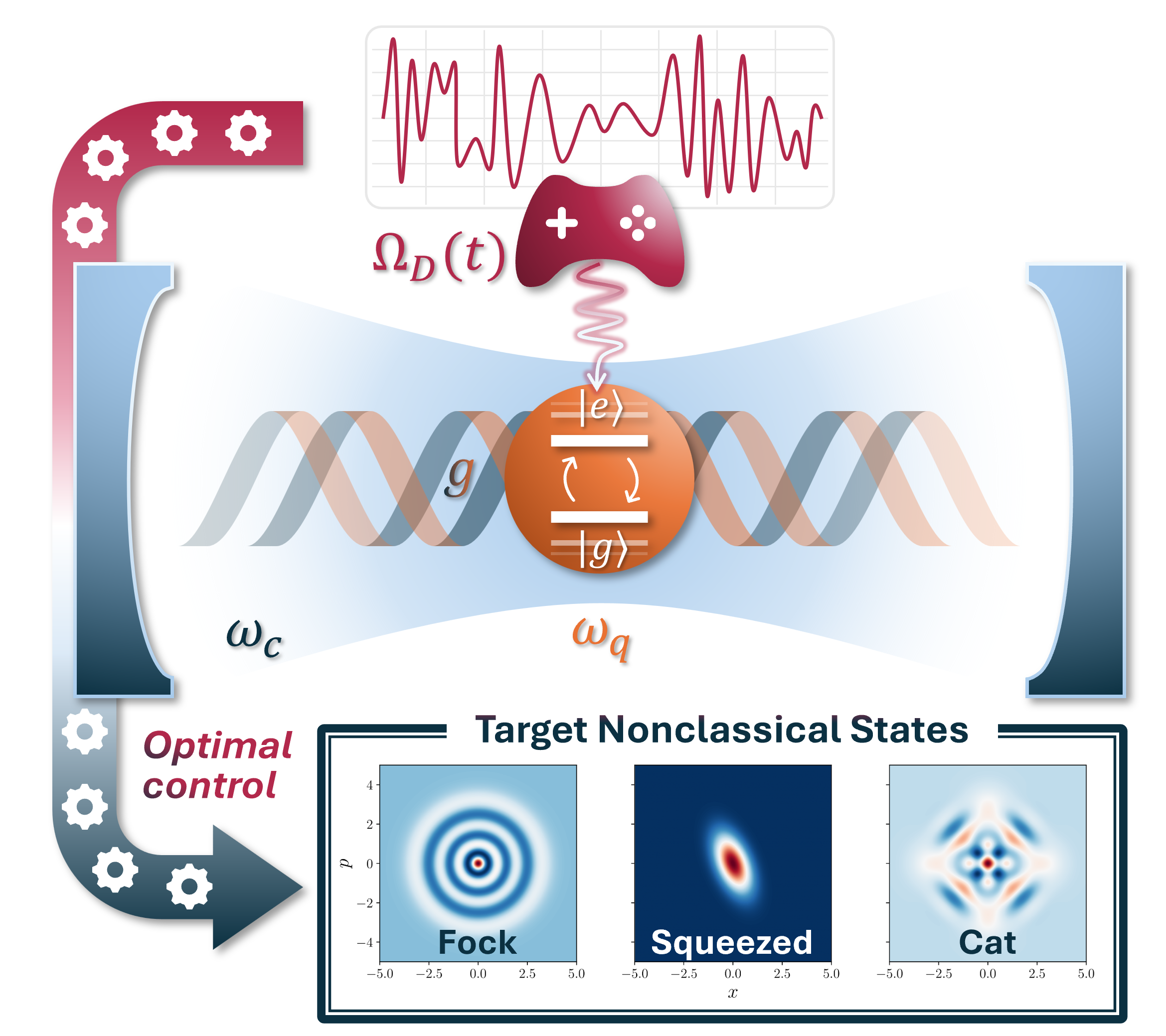}
    \caption{Schematic representation of a single-mode cavity of frequency $\omega_c$ coupled to a qubit of frequency $\omega_q$. An external drive $\Omega_D(t)$ modulates the qubit frequency, inducing dynamical Casimir photon generation in the USC regime. By optimally shaping the drive, target nonclassical cavity states can be generated deterministically from the quantum vacuum.
    }
    \label{fig:model}
\end{figure}

\textit{Model.}---We consider a high-quality single-mode electromagnetic cavity coupled to a two-level system (qubit) \cite{Harochebook, Beaudoin2011}, driven by a time-dependent control field acting on the qubit frequency (see Fig.~\ref{fig:model}).
The system is described by the Hamiltonian $H_S(t) = H_{R}+H_{D}(t)$, where the cavity-qubit interaction is described by the Rabi Hamiltonian 
\cite{Filipowicz:86,braak2011prl,xie2017jphysa}
(hereafter we set $\hbar = 1$) 
\begin{equation}
    H_{R} = \frac{\omega_q}{2}\sigma_z + \omega_c a^\dagger a + g \left(a^\dagger + a \right)\left(\sigma_+ + \sigma_-\right),
    \label{eq:H_Rabi}
\end{equation}
and the control Hamiltonian by
\begin{equation}
    H_{D}(t) = \frac{\Omega_{D}(t)}{2}\sigma_z
    \label{eq:H_Drive}
\end{equation}
Here, $a$ ($a^\dagger)$ is the bosonic annihilation (creation) operator, $\sigma_{+}=\ketbra{e}{g}$ ($\sigma_{-}=\sigma_+^\dagger$) is the qubit raising (lowering) operator, $\sigma_z=\dyad{e}-\dyad{g}$ is the Pauli-$z$ operator, $\omega_c$ ($\omega_q$) is the cavity (qubit) frequency, $g$ denotes the light--matter coupling strength, and $\Omega_D(t)$ is the external driving classical field. The state $\ket{g}$ ($\ket{e}$) is the ground (excited) state of the qubit. 
We focus on the USC regime ($g/\omega_c \gtrsim 0.1$) \cite{Beaudoin2011, Kockum2019, forndiaz2019}, where the counter-rotating terms in the cavity-qubit interaction, $a\sigma_- + a^\dagger\sigma_+$, are no longer negligible and contribute significantly to the system dynamics. These terms break the conservation of the total number of excitations, $n_{\rm ex} = a^\dagger a + \sigma_+\sigma_-$, allowing photons to be generated from the vacuum via the DCE 
\cite{Moore1970JMathPhys,Dodonov_2010,Wilson2011,RevModPhys.84.1,Hoeb2017,dodonov2020physics} \footnote{In this work, we consider the \textit{parametric} DCE, where photons are generated through the parametric amplification of vacuum fluctuations without moving  or changing the boundaries \cite{Dodonov_2010}}. While excitation number is not conserved, the parity
$\Pi = 
e^{i\pi n_{\rm ex} }$ is conserved by both the Rabi \cite{xie2017jphysa,braak2019symmetry} and the control Hamiltonians thus excitations are created/annihilated in pairs.

\textit{Optimal control strategy.}--- By modulating the instantaneous qubit frequency $\omega_q+\Omega_D(t)$, we harness the DCE to induce cavity excitations in a controlled way. Specifically, we initialize the system in the state $\ket{\psi_0} = \ket{0} \ket{g}$, where $\ket{0}$ denotes the cavity vacuum. In the absence of the counter-rotating terms, this state remains invariant under arbitrary modulations of the qubit frequency. 
By contrast, in the presence of counter-rotating terms, the modulation can drive the system away from the vacuum and steer it toward nonclassical target states of the cavity, such as Fock states (detailed below), as well as squeezed and Schr\"odinger-cat-state superpositions (see End Matter).
To this end, we employ the framework of quantum optimal control \cite{Koch2022} to shape the time-dependent modulation $\Omega_D(t)$ of the drive \eqref{eq:H_Drive}. Assuming the system evolves from time $t=0$ to a final time $t=T$, the control objective is to maximize the fidelity $F$ between the final evolved state and the chosen cavity target state. This is achieved by optimizing the control pulse to minimize the the cost function defined by the infidelity, $\mathcal{C} = 1 - F$ \footnote{Further details on the specific fidelity definitions used---which depend on the adopted optimal control technique---are provided in the Supplemental Material (SM).}.

We adopt a hybrid optimal control strategy that combines gradient-free and gradient-based methods. First, we use the Chopped RAndom Basis (CRAB) optimal control technique \cite{PhysRevA.84.022326,Muller2022}, a gradient-free method. 
Here, the control field is parametrized as a truncated Fourier expansion, 
\begin{equation}
    \Omega_D(t) = s(t)\sum_{k=1}^{N_f} \left[ A_k \sin(\omega_k t) + B_k \cos(\omega_k t) \right],
\end{equation}
where $\{A_k, B_k\}$ are variational parameters to be optimized, $N_f$ is the number of harmonics, and $\omega_k$ are randomly sampled frequencies. This parametrization naturally yields smooth control fields, while the envelope function $s(t)$ ensures smooth turn-on and turn-off of the external drive, enforcing $\Omega_D(0)=\Omega_D(T)=0$. Additionally, fixing the norm of the Fourier coefficients constrains the total control power, thereby restricting the CRAB optimization to physically realistic regimes.
The solution provided by CRAB is then fed as a high-quality initial guess for GRadient Ascent Pulse Engineering (GRAPE) \cite{KHANEJA2005296}, a gradient-based optimal control method. This allows GRAPE to refine the pulse profile to enhance the final fidelity. In GRAPE, the total evolution time $[0,T]$ is discretized into $N_t$ intervals of duration $\delta t = T/N_t$, and the control field is represented as a piecewise-constant function,
$\Omega_D(t) \rightarrow \{\Omega_{D}^{(1)}, \Omega_{D}^{(2)}, \ldots, \Omega_{D}^{(N_t)}\}.$
The control amplitudes $\{\Omega_{D}^{(k)}\}$ are iteratively updated using analytical gradients of the cost function, enforcing amplitude constraints to ensure compliance with the imposed power limitations. 
This GRAPE-based refinement step substantially enhances the fidelity. 
A final interpolation step is applied to smooth the control field, yielding a continuous and more experimentally feasible pulse with a negligible loss in fidelity.
Additional technical details on the optimization procedure are provided in the Supplemental material (SM).

Finally, inspired by counterdiabatic-driving methods \cite{PhysRevA.94.042132,campbell2017prl}, to quantify the energetic cost of the control fields for the nonclassical-state generation we compute the following state- and setup-independent cost functional
\begin{equation}
    C_2(t) := \int_0^t ds\, \|H_{D}(s)\|_2^2 
\label{eq:cost}
\end{equation}
for the drive Hamiltonian \eqref{eq:H_Drive}, where $\|X\|_2 = \sqrt{\Tr[X^\dagger X]}$ denotes the  
Frobenius norm \footnote{See \cite{tumbiolo2026prl} for the use of this metric beyond the counterdiabatic-driving regime.}.

\begin{figure*}[!ht]
    \centering
    \includegraphics[width=\linewidth]{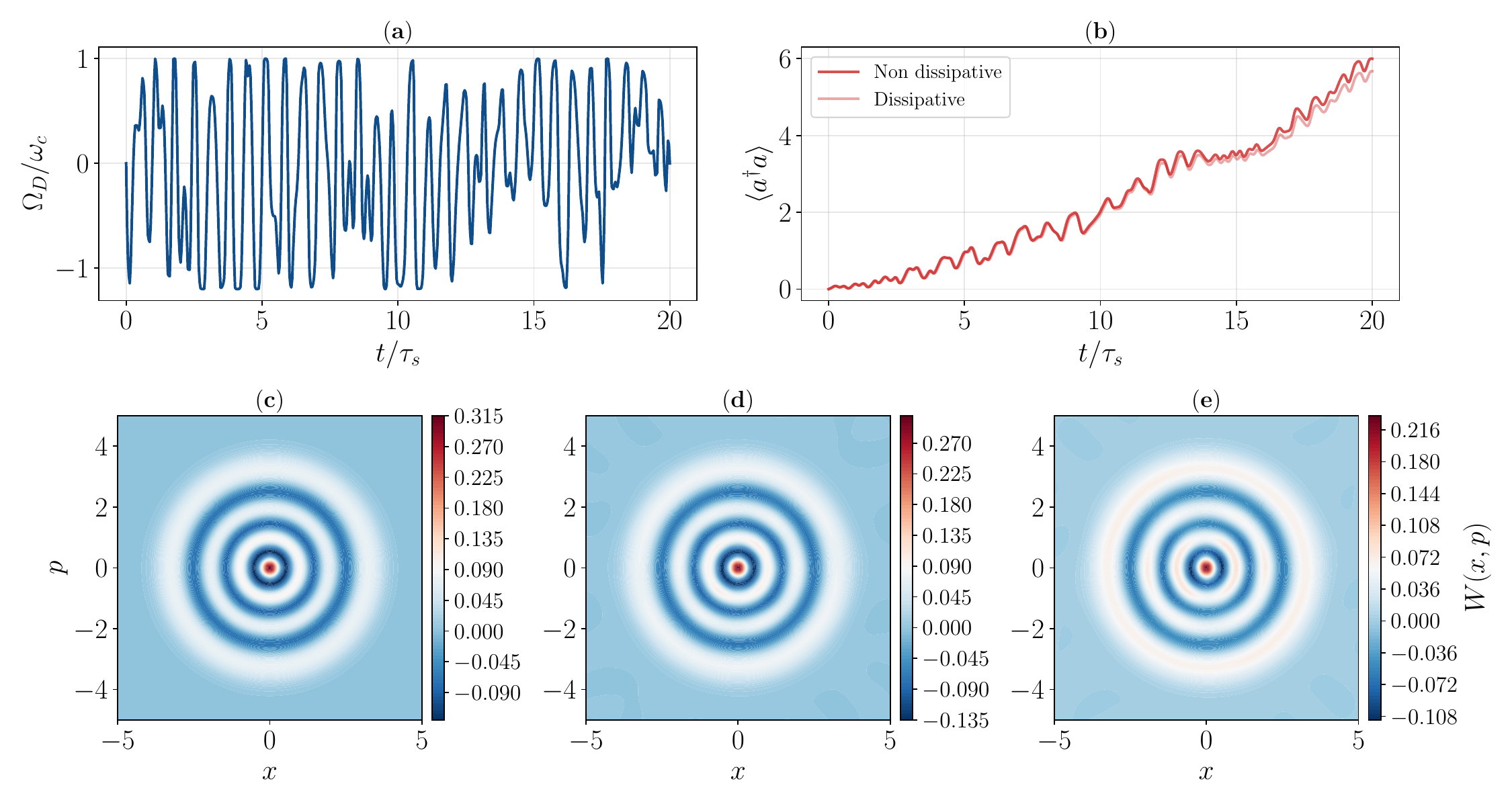}
    \caption{
    Optimal control strategy for the generation of the Fock state $\ket{n=6}$. (a) Optimized control drive $\Omega_D(t)/\omega_c$ obtained via CRAB optimization, GRAPE refinement, and final interpolation, and (b) corresponding cavity mean photon number $\langle a^\dagger a \rangle(t)$.
    (c--e) Wigner functions of the cavity field comparing (c) the ideal target state $\ket{n=6}$ with the state prepared by the optimized drive in (a) under (d) unitary dynamics (fidelity $F=0.995$) and (e) open-system dynamics (fidelity $F=0.901$). The phase-space distributions are plotted as a function of the field quadrature operators, defined as $x=(a+a^\dagger)/{\sqrt{2}}$ and $p=(a-a^\dagger)/{i\sqrt{2}}$.
    System and control parameters are $g{/\omega_c}=0.3$, $T=20\,\tau_s\equiv20\,\pi/2g$, $\omega_q{/\omega_c}=2$ (corresponding to the off-resonance regime $|\omega_c-\omega_q|\gg g$), and $N_f=20$. The number of time intervals for the GRAPE refinement is $N_t=300$.
     Environmental parameters are $\omega_B{/\omega_c}=10$, $\beta{\,\omega_c}=5$, and $\eta=10^{-4}$ (see section \textit{Robustness against noise} for details).
}
    \label{fig:drive}
\end{figure*}

\textit{Fock-state generation.}---
To demonstrate the effectiveness of our protocol, we tackle the generation of a highly nonclassical Fock state with $n=6$ as a first benchmark (see Fig. \ref{fig:drive}). Starting from the vacuum, the optimized control drive [panel (a)] steers the cavity population [panel (b)] to successfully prepare the desired $\ket{n=6}$ state with a fidelity of $F \approx 0.995$ over a total evolution time $T=20\,\tau_s$, where $\tau_s=\pi/2g$ sets the characteristic timescale \footnote{The timescale $\tau_s$ corresponds to the time required to perform half a vacuum Rabi oscillation in the resonant ($\omega_q=\omega_c$) Jaynes-Cummings limit [i.e., neglecting counter-rotating terms in the Rabi Hamiltonian (\ref{eq:H_Rabi})].}.
The excellent accuracy of the protocol is further confirmed by the final state's Wigner function, which faithfully reproduces the alternating, concentric interference rings characteristic of the ideal target state [panels (c) and (d)] \footnote{Depending on the specific experimental constraints, the fidelity $F \approx 0.995$ can be further enhanced up to $\approx 0.999$ by permitting larger control amplitudes (see SM).}.

\begin{figure}[!ht]
    \centering
    \includegraphics[width=\linewidth]{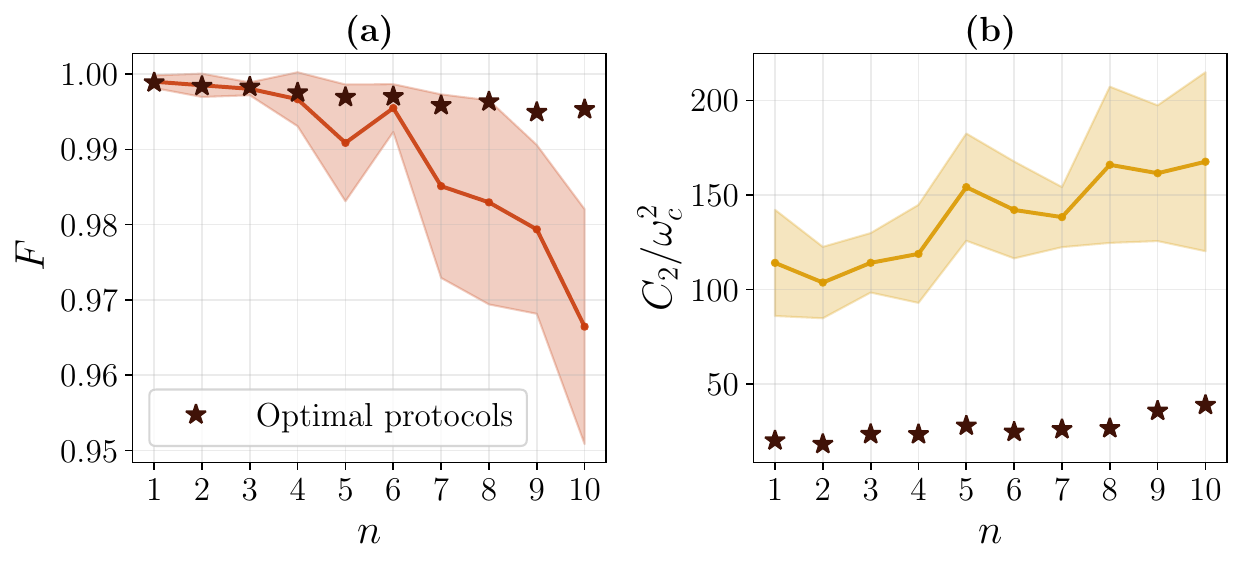}
    \caption{
    Performance scaling of Fock-state generation with the 
    target photon number $n$.  
    Average
    (a) fidelity, and
    (b) control cost \eqref{eq:cost} as a function of $n$.  
    Markers (mean value) and error bands (standard deviation) are obtained from 10 independent unconstrained CRAB optimizations with different random seeds; the star markers indicate the performance of the optimal protocols obtained by refining the best CRAB solutions through a subsequent GRAPE optimization followed by a final pulse-smoothing interpolation step. 
    The number of time intervals for the GRAPE refinement is $N_t=400$, 
    the other parameters are the same as those used in Fig.~\ref{fig:drive}.
    }
    \label{fig:fock_scaling}
\end{figure}

To evaluate how the protocol performance scales with increasing photon number, we investigate the generation of Fock states over the range $1 \leq n \leq 10$. 
We first consider bare CRAB optimization, without GRAPE refinement or constraints on the control amplitudes. The protocol performance is assessed in terms of the final-state fidelity and the control cost defined in Eq.~\eqref{eq:cost}. For each $n$, both quantities are averaged over 10 independent CRAB optimizations with different random seeds (see Fig.~\ref{fig:fock_scaling}).
The average fidelity is consistently high, $F \gtrsim 0.96$, across all target states [panel (a)], confirming the robustness of the control strategy beyond the low-excitation regime. As $n$ increases, however, the average fidelity gradually decreases, while its dispersion widens, indicating an increasingly intricate control landscape for higher photon numbers. 
Consistently, the cost [panel (b)] mildly increases with $n$, reflecting the greater energetic resources required to target highly excited Fock states.
 
Figure~\ref{fig:fock_scaling} also presents the results obtained by taking, for each $n$, the highest-fidelity CRAB solution as an initial guess for a constrained GRAPE refinement, followed by pulse smoothing.
Notably, the refined fidelity systematically outperforms the bare CRAB average, particularly in the large-$n$ regime. Crucially, the control cost is drastically reduced, proving that amplitude bounds effectively guide the protocol toward far more energetically efficient solutions.

\textit{Robustness against noise.}---
We assess the robustness of the optimized control strategy from Fig. \ref{fig:drive} against both classical noise affecting the control field and quantum noise induced by coupling to an external environment. 

\begin{figure}
    \centering
    \includegraphics[width=1\linewidth]{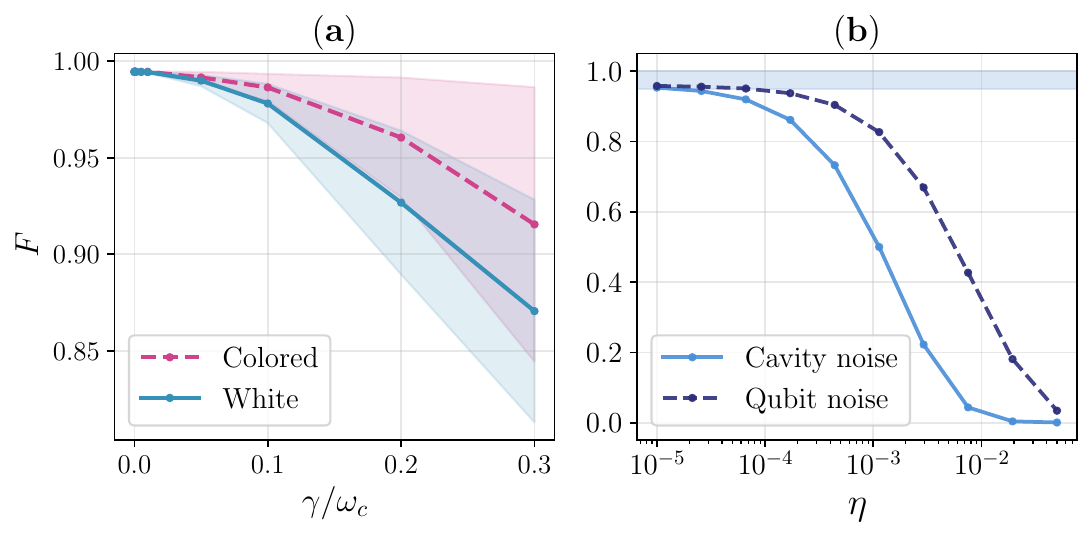}
    \caption{
    Robustness of Fock-state generation against classical and quantum noise. (a) Classical control noise: Fidelity as a function of the dimensionless noise amplitude $\gamma/\omega_c$  
    for two additive noise models (white noise, and colored noise with finite correlation time $\tau_c=\tau_s/3$). Markers (mean value) and error bands (standard deviation) are obtained from 100 independent realizations.
    (b) Quantum noise: Fidelity as a function of the system--bath coupling strength $\eta$ for Lindblad operators acting on the cavity (solid curve) and on the qubit (dashed curve). 
    The high-fidelity region ($0.95 \leq F \leq 1$) is highlighted for reference.
    System parameters are the same as those used for the optimized control pulse in Fig.~\ref{fig:drive}.
    Environmental parameters (see SM) are $\omega_B{/\omega_c} = 10$ and $\beta{\,\omega_c} = 5$.}
    \label{fig:classical_noise}
\end{figure}

As \textit{classical control noise}, we consider two representative additive noises [$\Omega_D(t) \to \Omega_D(t) + \delta \Omega_D(t)$]: (i) white noise $\delta \Omega_D(t) = \gamma \xi(t)$, described as an uncorrelated Gaussian process $\xi(t)$ with zero mean and unit variance, and (ii) colored noise $\delta \Omega_D(t)=\gamma \tilde{\xi}(t)$, obtained by filtering white noise through a Gaussian kernel $G_{\tau_c}$ with correlation time $\tau_c$, yielding a correlated process $\tilde{\xi}(t) = (\xi * G_{\tau_c})(t)$. The noise amplitude is parametrized by $\gamma$. These additive noises capture broadband fluctuations and finite-bandwidth noise, respectively. 
For each $\gamma$, we average the fidelity over $100$ independent stochastic realizations of the noise, and show it against the noise amplitude in  Fig.~\ref{fig:classical_noise}(a). The protocol remains highly robust against both noise models, maintaining fidelities $F \gtrsim 0.95$ up to $\gamma \sim 0.1$ for white noise, and up to $\gamma \sim 0.2$ for colored noise. 
The relatively higher robustness against colored noise can be understood from the broadband nature of white noise, whose high-frequency components effectively disrupt the dynamics.

Turning to \textit{quantum noise}, we couple the the cavity-qubit system to a thermal bosonic environment at inverse temperature $\beta$. The environment is modeled as a bath of harmonic oscillators with an Ohmic spectral density with exponential cutoff, $J(\omega) = \eta |\omega| e^{-|\omega| / \omega_{B}}$, where $\eta$ sets the system--bath coupling strength and $\omega_{B}$ is the bath cutoff frequency. The reduced system dynamics are described within the Gorini--Kossakowski--Lindblad--Sudarshan (GKLS) formalism~\cite{breuer2002theory,lindblad1976generators,gorini1976completely}, valid under the assumptions of weak system--bath coupling and Markovian dynamics. The evolution of the system density matrix $\rho_S(t)$ is governed by 
the GKLS master equation 
\begin{align}
&\frac{d}{dt}\rho_S(t) = -i[H_S(t),\rho_S(t)] 
\nonumber\\&+ 
\sum_\omega \left( L_\omega(t) \rho_S(t) L_\omega^\dagger(t) - \frac{1}{2}\{L_\omega^\dagger(t) L_\omega(t),\rho_S(t)\} \right),
    \label{eq:GKSL_me}
\end{align}
where the Lindblad operators $L_\omega(t)$ account for the dissipative channels,
acting either on the cavity or on the qubit (see SM for details on the microscopic model). 
Figure~\ref{fig:classical_noise}(b) characterizes the robustness of the optimal $\ket{n=6}$ Fock-state generation from Fig. \ref{fig:drive} as a function of the system--bath coupling strength $\eta$. The fidelity remains remarkably high ($F \gtrsim 0.9$) up to $\eta \lesssim 10^{-4}$ (with qubit noise exhibiting consistently higher fidelity values compared to cavity noise),
before rapidly decreasing at larger noise amplitudes. At this representative threshold $\eta = 10^{-4}$, the state remains highly faithful to the target Fock state: as shown by the Wigner function in Fig.~\ref{fig:drive}(e) [see also panel (b), where the mean photon number obtained in the dissipative case closely tracks the corresponding nondissipative evolution], the protocol yields a fidelity of $F = 0.901$, compared with the ideal value of $F = 0.995$ obtained under unitary, noise-free evolution. 

\textit{Conclusions and discussion}.---We have shown that the DCE in a cavity--qubit system operating in the USC regime can be harnessed for the high-fidelity generation of nonclassical cavity states. By employing optimal-control techniques, we design driving protocols that modulate the qubit frequency and convert vacuum fluctuations into controllable excitations, thereby enabling the preparation of target states. Furthermore, our hybrid optimization strategy naturally accommodates constraints on the control amplitudes, limiting the overall control power and facilitating adaptation to experimental requirements. 

The generation of Fock states $\ket{n}$ is remarkably effective, yielding fidelities exceeding $0.99$ up to $n=10$, while remaining highly robust against both classical control noise and quantum dissipation. 
Importantly, the approach extends beyond Fock-state preparation to a broad class of nonclassical states, as illustrated by squeezed-vacuum states and Schr\"{o}dinger-cat-state superpositions (see End Matter). 

The proposed protocol represents a promising route toward the realization of ultrafast, robust, high-fidelity nonclassical-state generation in circuit-QED platforms operating in the USC regime \cite{yoshihara2017natphys,yoshihara2017pra}. 
The optimized control pulses can be implemented using currently available microwave-control technology \cite{deng2015prl},
while the open-system simulations here discussed employ system/bath parameters representative of contemporary superconducting devices \cite{Niemczyk2010,Fink2008}.
Moreover, the USC regime enables state-preparation times substantially shorter than those achieved in weak-coupling architectures. Additional details and quantitative benchmarks are provided in the SM. 

Overall, our findings establish the DCE as a resource---rather than a limitation---for quantum-state engineering. The impact of the proposed protocol is expected to be particularly significant in applications where state preparation must be deterministic rather than heralded \cite{Yoshikawa2013prx}. 
These include quantum computing, ranging from quantum gate operations \cite{kurman2026prx} and continuous-variable encodings \cite{lau2016PhysRevLett} to quantum error correction \cite{michael2016prx,Ofek2016nature,lee2024prxquantum,xu2023npjquantuminfo}, as well as quantum communication \cite{paparelle2025optexpress,Reiserer2015RevModPhys} and quantum metrology \cite{schnabel2017physrep}, where nonclassical states must be generated on demand and, in many cases, at high repetition rates.

More broadly, extending the optimal-control framework developed here to multi-qubit cavity architectures may open new avenues for scalable quantum-state engineering. The large Hilbert spaces and intrinsic nonlinear dynamics of such systems also make them attractive candidates for quantum reservoir computing \cite{MartinezPena2023,Sannia2024,Zhu2025,Das2026}.

\textit{Acknowledgments.}---
M.C. and G.B. acknowledge support from INFN through the project “QUANTUM”.
L.R. acknowledges support from University of Pavia through the project ``Termodinamica di precisione per sistemi aperti quantistici'', funded within the ``Fondo Ricerca e Giovani 2024'' programme.
G.G. kindly acknowledges support from the Ministero dell’Università e della Ricerca (MUR) under the “Rita Levi-Montalcini” grant.
G.G. and L.R. also acknowledge support from INFN through the project ``BELL''.
L.G. acknowledges support from the PNRR MUR project PE0000023-NQSTI, ``National Quantum Science and Technology Institute''.
G.A.F. acknowledges support from the ICSC - Centro Nazionale di Ricerca in High-Performance Computing, Big Data and Quantum Computing.
L.G. and G.A.F. acknowledge support from PRIN 2022 ``SuperNISQ'' and from the University of Catania, Piano Incentivi Ricerca di Ateneo 2024-26, project TCMQI.

\bibliography{references}

\begin{thebibliography}{26}%
\makeatletter
\providecommand \@ifxundefined [1]{%
 \@ifx{#1\undefined}
}%
\providecommand \@ifnum [1]{%
 \ifnum #1\expandafter \@firstoftwo
 \else \expandafter \@secondoftwo
 \fi
}%
\providecommand \@ifx [1]{%
 \ifx #1\expandafter \@firstoftwo
 \else \expandafter \@secondoftwo
 \fi
}%
\providecommand \natexlab [1]{#1}%
\providecommand \enquote  [1]{``#1''}%
\providecommand \bibnamefont  [1]{#1}%
\providecommand \bibfnamefont [1]{#1}%
\providecommand \citenamefont [1]{#1}%
\providecommand \href@noop [0]{\@secondoftwo}%
\providecommand \href [0]{\begingroup \@sanitize@url \@href}%
\providecommand \@href[1]{\@@startlink{#1}\@@href}%
\providecommand \@@href[1]{\endgroup#1\@@endlink}%
\providecommand \@sanitize@url [0]{\catcode `\\12\catcode `\$12\catcode `\&12\catcode `\#12\catcode `\^12\catcode `\_12\catcode `\%12\relax}%
\providecommand \@@startlink[1]{}%
\providecommand \@@endlink[0]{}%
\providecommand \url  [0]{\begingroup\@sanitize@url \@url }%
\providecommand \@url [1]{\endgroup\@href {#1}{\urlprefix }}%
\providecommand \urlprefix  [0]{URL }%
\providecommand \Eprint [0]{\href }%
\providecommand \doibase [0]{https://doi.org/}%
\providecommand \selectlanguage [0]{\@gobble}%
\providecommand \bibinfo  [0]{\@secondoftwo}%
\providecommand \bibfield  [0]{\@secondoftwo}%
\providecommand \translation [1]{[#1]}%
\providecommand \BibitemOpen [0]{}%
\providecommand \bibitemStop [0]{}%
\providecommand \bibitemNoStop [0]{.\EOS\space}%
\providecommand \EOS [0]{\spacefactor3000\relax}%
\providecommand \BibitemShut  [1]{\csname bibitem#1\endcsname}%
\let\auto@bib@innerbib\@empty
\bibitem [{\citenamefont {Caneva}\ \emph {et~al.}(2011)\citenamefont {Caneva}, \citenamefont {Calarco},\ and\ \citenamefont {Montangero}}]{PhysRevA.84.022326}%
  \BibitemOpen
  \bibfield  {author} {\bibinfo {author} {\bibfnamefont {T.}~\bibnamefont {Caneva}}, \bibinfo {author} {\bibfnamefont {T.}~\bibnamefont {Calarco}},\ and\ \bibinfo {author} {\bibfnamefont {S.}~\bibnamefont {Montangero}},\ }\bibfield  {title} {\bibinfo {title} {Chopped random-basis quantum optimization},\ }\href {https://doi.org/10.1103/PhysRevA.84.022326} {\bibfield  {journal} {\bibinfo  {journal} {Phys. Rev. A}\ }\textbf {\bibinfo {volume} {84}},\ \bibinfo {pages} {022326} (\bibinfo {year} {2011})}\BibitemShut {NoStop}%
\bibitem [{\citenamefont {Müller}\ \emph {et~al.}(2022)\citenamefont {Müller}, \citenamefont {Said}, \citenamefont {Jelezko}, \citenamefont {Calarco},\ and\ \citenamefont {Montangero}}]{Muller2022}%
  \BibitemOpen
  \bibfield  {author} {\bibinfo {author} {\bibfnamefont {M.~M.}\ \bibnamefont {Müller}}, \bibinfo {author} {\bibfnamefont {R.~S.}\ \bibnamefont {Said}}, \bibinfo {author} {\bibfnamefont {F.}~\bibnamefont {Jelezko}}, \bibinfo {author} {\bibfnamefont {T.}~\bibnamefont {Calarco}},\ and\ \bibinfo {author} {\bibfnamefont {S.}~\bibnamefont {Montangero}},\ }\bibfield  {title} {\bibinfo {title} {One decade of quantum optimal control in the chopped random basis},\ }\href {https://doi.org/10.1088/1361-6633/ac723c} {\bibfield  {journal} {\bibinfo  {journal} {Reports on Progress in Physics}\ }\textbf {\bibinfo {volume} {85}},\ \bibinfo {pages} {076001} (\bibinfo {year} {2022})}\BibitemShut {NoStop}%
\bibitem [{\citenamefont {Khaneja}\ \emph {et~al.}(2005)\citenamefont {Khaneja}, \citenamefont {Reiss}, \citenamefont {Kehlet}, \citenamefont {Schulte-Herbrüggen},\ and\ \citenamefont {Glaser}}]{KHANEJA2005296}%
  \BibitemOpen
  \bibfield  {author} {\bibinfo {author} {\bibfnamefont {N.}~\bibnamefont {Khaneja}}, \bibinfo {author} {\bibfnamefont {T.}~\bibnamefont {Reiss}}, \bibinfo {author} {\bibfnamefont {C.}~\bibnamefont {Kehlet}}, \bibinfo {author} {\bibfnamefont {T.}~\bibnamefont {Schulte-Herbrüggen}},\ and\ \bibinfo {author} {\bibfnamefont {S.~J.}\ \bibnamefont {Glaser}},\ }\bibfield  {title} {\bibinfo {title} {Optimal control of coupled spin dynamics: design of nmr pulse sequences by gradient ascent algorithms},\ }\href {https://doi.org/https://doi.org/10.1016/j.jmr.2004.11.004} {\bibfield  {journal} {\bibinfo  {journal} {Journal of Magnetic Resonance}\ }\textbf {\bibinfo {volume} {172}},\ \bibinfo {pages} {296} (\bibinfo {year} {2005})}\BibitemShut {NoStop}%
\bibitem [{\citenamefont {Johansson}\ \emph {et~al.}(2012)\citenamefont {Johansson}, \citenamefont {Nation},\ and\ \citenamefont {Nori}}]{Johansson2012}%
  \BibitemOpen
  \bibfield  {author} {\bibinfo {author} {\bibfnamefont {J.~R.}\ \bibnamefont {Johansson}}, \bibinfo {author} {\bibfnamefont {P.~D.}\ \bibnamefont {Nation}},\ and\ \bibinfo {author} {\bibfnamefont {F.}~\bibnamefont {Nori}},\ }\bibfield  {title} {\bibinfo {title} {{QuTiP: An open-source Python framework for the dynamics of open quantum systems}},\ }\href {https://doi.org/10.1016/j.cpc.2012.02.021} {\bibfield  {journal} {\bibinfo  {journal} {Computer Physics Communications}\ }\textbf {\bibinfo {volume} {183}},\ \bibinfo {pages} {1760} (\bibinfo {year} {2012})}\BibitemShut {NoStop}%
\bibitem [{\citenamefont {Johansson}\ \emph {et~al.}(2013)\citenamefont {Johansson}, \citenamefont {Nation},\ and\ \citenamefont {Nori}}]{Johansson2013}%
  \BibitemOpen
  \bibfield  {author} {\bibinfo {author} {\bibfnamefont {J.~R.}\ \bibnamefont {Johansson}}, \bibinfo {author} {\bibfnamefont {P.~D.}\ \bibnamefont {Nation}},\ and\ \bibinfo {author} {\bibfnamefont {F.}~\bibnamefont {Nori}},\ }\bibfield  {title} {\bibinfo {title} {{QuTiP 2: A Python framework for the dynamics of open quantum systems}},\ }\href {https://doi.org/10.1016/j.cpc.2012.11.019} {\bibfield  {journal} {\bibinfo  {journal} {Computer Physics Communications}\ }\textbf {\bibinfo {volume} {184}},\ \bibinfo {pages} {1234} (\bibinfo {year} {2013})}\BibitemShut {NoStop}%
\bibitem [{\citenamefont {Gao}\ and\ \citenamefont {Han}(2012)}]{Gao2012}%
  \BibitemOpen
  \bibfield  {author} {\bibinfo {author} {\bibfnamefont {F.}~\bibnamefont {Gao}}\ and\ \bibinfo {author} {\bibfnamefont {L.}~\bibnamefont {Han}},\ }\bibfield  {title} {\bibinfo {title} {Implementing the nelder-mead simplex algorithm with adaptive parameters},\ }\href {https://doi.org/10.1007/s10589-010-9329-3} {\bibfield  {journal} {\bibinfo  {journal} {Computational Optimization and Applications}\ }\textbf {\bibinfo {volume} {51}},\ \bibinfo {pages} {259} (\bibinfo {year} {2012})}\BibitemShut {NoStop}%
\bibitem [{\citenamefont {Virtanen}\ \emph {et~al.}(2020)\citenamefont {Virtanen}, \citenamefont {Gommers}, \citenamefont {Oliphant}, \citenamefont {Haberland}, \citenamefont {Reddy}, \citenamefont {Cournapeau}, \citenamefont {Burovski}, \citenamefont {Peterson}, \citenamefont {Weckesser}, \citenamefont {Bright}, \citenamefont {van~der Walt}, \citenamefont {Brett}, \citenamefont {Wilson}, \citenamefont {Millman}, \citenamefont {Mayorov}, \citenamefont {Nelson}, \citenamefont {Jones}, \citenamefont {Kern}, \citenamefont {Larson}, \citenamefont {Carey}, \citenamefont {Polat}, \citenamefont {Feng}, \citenamefont {Moore}, \citenamefont {VanderPlas}, \citenamefont {Laxalde}, \citenamefont {Perktold}, \citenamefont {Cimrman}, \citenamefont {Henriksen}, \citenamefont {Quintero}, \citenamefont {Harris}, \citenamefont {Archibald}, \citenamefont {Ribeiro}, \citenamefont {Pedregosa}, \citenamefont {van Mulbregt},\ and\ \citenamefont {the SciPy 1.0~Contributors}}]{2020SciPyNMeth}%
  \BibitemOpen
  \bibfield  {author} {\bibinfo {author} {\bibfnamefont {P.}~\bibnamefont {Virtanen}}, \bibinfo {author} {\bibfnamefont {R.}~\bibnamefont {Gommers}}, \bibinfo {author} {\bibfnamefont {T.~E.}\ \bibnamefont {Oliphant}}, \bibinfo {author} {\bibfnamefont {M.}~\bibnamefont {Haberland}}, \bibinfo {author} {\bibfnamefont {T.}~\bibnamefont {Reddy}}, \bibinfo {author} {\bibfnamefont {D.}~\bibnamefont {Cournapeau}}, \bibinfo {author} {\bibfnamefont {E.}~\bibnamefont {Burovski}}, \bibinfo {author} {\bibfnamefont {P.}~\bibnamefont {Peterson}}, \bibinfo {author} {\bibfnamefont {W.}~\bibnamefont {Weckesser}}, \bibinfo {author} {\bibfnamefont {J.}~\bibnamefont {Bright}}, \bibinfo {author} {\bibfnamefont {S.~J.}\ \bibnamefont {van~der Walt}}, \bibinfo {author} {\bibfnamefont {M.}~\bibnamefont {Brett}}, \bibinfo {author} {\bibfnamefont {J.}~\bibnamefont {Wilson}}, \bibinfo {author} {\bibfnamefont {K.~J.}\ \bibnamefont {Millman}}, \bibinfo {author} {\bibfnamefont {N.}~\bibnamefont {Mayorov}}, \bibinfo {author} {\bibfnamefont
  {A.~R.~J.}\ \bibnamefont {Nelson}}, \bibinfo {author} {\bibfnamefont {E.}~\bibnamefont {Jones}}, \bibinfo {author} {\bibfnamefont {R.}~\bibnamefont {Kern}}, \bibinfo {author} {\bibfnamefont {E.}~\bibnamefont {Larson}}, \bibinfo {author} {\bibfnamefont {C.~J.}\ \bibnamefont {Carey}}, \bibinfo {author} {\bibfnamefont {{\.{I}}.}~\bibnamefont {Polat}}, \bibinfo {author} {\bibfnamefont {Y.}~\bibnamefont {Feng}}, \bibinfo {author} {\bibfnamefont {E.~W.}\ \bibnamefont {Moore}}, \bibinfo {author} {\bibfnamefont {J.}~\bibnamefont {VanderPlas}}, \bibinfo {author} {\bibfnamefont {D.}~\bibnamefont {Laxalde}}, \bibinfo {author} {\bibfnamefont {J.}~\bibnamefont {Perktold}}, \bibinfo {author} {\bibfnamefont {R.}~\bibnamefont {Cimrman}}, \bibinfo {author} {\bibfnamefont {I.}~\bibnamefont {Henriksen}}, \bibinfo {author} {\bibfnamefont {E.~A.}\ \bibnamefont {Quintero}}, \bibinfo {author} {\bibfnamefont {C.~R.}\ \bibnamefont {Harris}}, \bibinfo {author} {\bibfnamefont {A.~M.}\ \bibnamefont {Archibald}}, \bibinfo {author}
  {\bibfnamefont {A.~H.}\ \bibnamefont {Ribeiro}}, \bibinfo {author} {\bibfnamefont {F.}~\bibnamefont {Pedregosa}}, \bibinfo {author} {\bibfnamefont {P.}~\bibnamefont {van Mulbregt}},\ and\ \bibinfo {author} {\bibnamefont {the SciPy 1.0~Contributors}},\ }\bibfield  {title} {\bibinfo {title} {Scipy 1.0: Fundamental algorithms for scientific computing in python},\ }\href {https://doi.org/10.1038/s41592-019-0686-2} {\bibfield  {journal} {\bibinfo  {journal} {Nature Methods}\ }\textbf {\bibinfo {volume} {17}},\ \bibinfo {pages} {261} (\bibinfo {year} {2020})}\BibitemShut {NoStop}%
\bibitem [{\citenamefont {Byrd}\ \emph {et~al.}(1995)\citenamefont {Byrd}, \citenamefont {Lu}, \citenamefont {Nocedal},\ and\ \citenamefont {Zhu}}]{LBFGSB1995}%
  \BibitemOpen
  \bibfield  {author} {\bibinfo {author} {\bibfnamefont {R.~H.}\ \bibnamefont {Byrd}}, \bibinfo {author} {\bibfnamefont {P.}~\bibnamefont {Lu}}, \bibinfo {author} {\bibfnamefont {J.}~\bibnamefont {Nocedal}},\ and\ \bibinfo {author} {\bibfnamefont {C.}~\bibnamefont {Zhu}},\ }\bibfield  {title} {\bibinfo {title} {A limited memory algorithm for bound constrained optimization},\ }\href {https://doi.org/10.1137/0916069} {\bibfield  {journal} {\bibinfo  {journal} {SIAM Journal on Scientific Computing}\ }\textbf {\bibinfo {volume} {16}},\ \bibinfo {pages} {1190} (\bibinfo {year} {1995})}\BibitemShut {NoStop}%
\bibitem [{\citenamefont {Scully}\ and\ \citenamefont {Zubairy}(1997)}]{Scully_Zubairy_1997}%
  \BibitemOpen
  \bibfield  {author} {\bibinfo {author} {\bibfnamefont {M.~O.}\ \bibnamefont {Scully}}\ and\ \bibinfo {author} {\bibfnamefont {M.~S.}\ \bibnamefont {Zubairy}},\ }\href@noop {} {\emph {\bibinfo {title} {Quantum Optics}}}\ (\bibinfo  {publisher} {Cambridge University Press},\ \bibinfo {year} {1997})\BibitemShut {NoStop}%
\bibitem [{\citenamefont {Carrega}\ \emph {et~al.}(2024)\citenamefont {Carrega}, \citenamefont {Razzoli}, \citenamefont {Erdman}, \citenamefont {Cavaliere}, \citenamefont {Benenti},\ and\ \citenamefont {Sassetti}}]{10.1116/5.0190340}%
  \BibitemOpen
  \bibfield  {author} {\bibinfo {author} {\bibfnamefont {M.}~\bibnamefont {Carrega}}, \bibinfo {author} {\bibfnamefont {L.}~\bibnamefont {Razzoli}}, \bibinfo {author} {\bibfnamefont {P.~A.}\ \bibnamefont {Erdman}}, \bibinfo {author} {\bibfnamefont {F.}~\bibnamefont {Cavaliere}}, \bibinfo {author} {\bibfnamefont {G.}~\bibnamefont {Benenti}},\ and\ \bibinfo {author} {\bibfnamefont {M.}~\bibnamefont {Sassetti}},\ }\bibfield  {title} {\bibinfo {title} {Dissipation-induced collective advantage of a quantum thermal machine},\ }\href {https://doi.org/10.1116/5.0190340} {\bibfield  {journal} {\bibinfo  {journal} {AVS Quantum Science}\ }\textbf {\bibinfo {volume} {6}},\ \bibinfo {pages} {025001} (\bibinfo {year} {2024})}\BibitemShut {NoStop}%
\bibitem [{\citenamefont {Breuer}\ and\ \citenamefont {Petruccione}(2007)}]{breuer2002theory}%
  \BibitemOpen
  \bibfield  {author} {\bibinfo {author} {\bibfnamefont {H.-P.}\ \bibnamefont {Breuer}}\ and\ \bibinfo {author} {\bibfnamefont {F.}~\bibnamefont {Petruccione}},\ }\href {https://doi.org/10.1093/acprof:oso/9780199213900.001.0001} {\emph {\bibinfo {title} {The Theory of Open Quantum Systems}}}\ (\bibinfo  {publisher} {Oxford University Press},\ \bibinfo {year} {2007})\BibitemShut {NoStop}%
\bibitem [{\citenamefont {Manzano}(2020)}]{Manzano}%
  \BibitemOpen
  \bibfield  {author} {\bibinfo {author} {\bibfnamefont {D.}~\bibnamefont {Manzano}},\ }\bibfield  {title} {\bibinfo {title} {A short introduction to the {L}indblad master equation},\ }\href {https://doi.org/10.1063/1.5115323} {\bibfield  {journal} {\bibinfo  {journal} {AIP Advances}\ }\textbf {\bibinfo {volume} {10}},\ \bibinfo {pages} {025106} (\bibinfo {year} {2020})}\BibitemShut {NoStop}%
\bibitem [{\citenamefont {Campaioli}\ \emph {et~al.}(2024)\citenamefont {Campaioli}, \citenamefont {Cole},\ and\ \citenamefont {Hapuarachchi}}]{Campaioli}%
  \BibitemOpen
  \bibfield  {author} {\bibinfo {author} {\bibfnamefont {F.}~\bibnamefont {Campaioli}}, \bibinfo {author} {\bibfnamefont {J.~H.}\ \bibnamefont {Cole}},\ and\ \bibinfo {author} {\bibfnamefont {H.}~\bibnamefont {Hapuarachchi}},\ }\bibfield  {title} {\bibinfo {title} {Quantum master equations: Tips and tricks for quantum optics, quantum computing, and beyond},\ }\href {https://doi.org/10.1103/PRXQuantum.5.020202} {\bibfield  {journal} {\bibinfo  {journal} {PRX Quantum}\ }\textbf {\bibinfo {volume} {5}},\ \bibinfo {pages} {020202} (\bibinfo {year} {2024})}\BibitemShut {NoStop}%
\bibitem [{\citenamefont {Vacchini}(2024)}]{vacchini2024open}%
  \BibitemOpen
  \bibfield  {author} {\bibinfo {author} {\bibfnamefont {B.}~\bibnamefont {Vacchini}},\ }\href {https://doi.org/10.1007/978-3-031-58218-9} {\emph {\bibinfo {title} {Open Quantum Systems: Foundations and Theory}}},\ \bibinfo {edition} {1st}\ ed.,\ Graduate Texts in Physics\ (\bibinfo  {publisher} {Springer},\ \bibinfo {address} {Cham},\ \bibinfo {year} {2024})\BibitemShut {NoStop}%
\bibitem [{\citenamefont {Kossakowski}(1972)}]{kossakowski1972nonhamiltonian}%
  \BibitemOpen
  \bibfield  {author} {\bibinfo {author} {\bibfnamefont {A.}~\bibnamefont {Kossakowski}},\ }\bibfield  {title} {\bibinfo {title} {On quantum statistical mechanics of non-{H}amiltonian systems},\ }\href {https://doi.org/10.1016/0034-4877(72)90010-9} {\bibfield  {journal} {\bibinfo  {journal} {Reports on Mathematical Physics}\ }\textbf {\bibinfo {volume} {3}},\ \bibinfo {pages} {247} (\bibinfo {year} {1972})}\BibitemShut {NoStop}%
\bibitem [{\citenamefont {Lindblad}(1976)}]{lindblad1976generators}%
  \BibitemOpen
  \bibfield  {author} {\bibinfo {author} {\bibfnamefont {G.}~\bibnamefont {Lindblad}},\ }\bibfield  {title} {\bibinfo {title} {On the generators of quantum dynamical semigroups},\ }\href {https://doi.org/10.1007/BF01608499} {\bibfield  {journal} {\bibinfo  {journal} {Communications in Mathematical Physics}\ }\textbf {\bibinfo {volume} {48}},\ \bibinfo {pages} {119} (\bibinfo {year} {1976})}\BibitemShut {NoStop}%
\bibitem [{\citenamefont {Gorini}\ \emph {et~al.}(1976)\citenamefont {Gorini}, \citenamefont {Kossakowski},\ and\ \citenamefont {Sudarshan}}]{gorini1976completely}%
  \BibitemOpen
  \bibfield  {author} {\bibinfo {author} {\bibfnamefont {V.}~\bibnamefont {Gorini}}, \bibinfo {author} {\bibfnamefont {A.}~\bibnamefont {Kossakowski}},\ and\ \bibinfo {author} {\bibfnamefont {E.~C.~G.}\ \bibnamefont {Sudarshan}},\ }\bibfield  {title} {\bibinfo {title} {Completely positive dynamical semigroups of {$N$}‐level systems},\ }\href {https://doi.org/10.1063/1.522979} {\bibfield  {journal} {\bibinfo  {journal} {Journal of Mathematical Physics}\ }\textbf {\bibinfo {volume} {17}},\ \bibinfo {pages} {821} (\bibinfo {year} {1976})}\BibitemShut {NoStop}%
\bibitem [{\citenamefont {Yoshihara}\ \emph {et~al.}(2017{\natexlab{a}})\citenamefont {Yoshihara}, \citenamefont {Fuse}, \citenamefont {Ashhab}, \citenamefont {Kakuyanagi}, \citenamefont {Saito},\ and\ \citenamefont {Semba}}]{yoshihara2017natphys}%
  \BibitemOpen
  \bibfield  {author} {\bibinfo {author} {\bibfnamefont {F.}~\bibnamefont {Yoshihara}}, \bibinfo {author} {\bibfnamefont {T.}~\bibnamefont {Fuse}}, \bibinfo {author} {\bibfnamefont {S.}~\bibnamefont {Ashhab}}, \bibinfo {author} {\bibfnamefont {K.}~\bibnamefont {Kakuyanagi}}, \bibinfo {author} {\bibfnamefont {S.}~\bibnamefont {Saito}},\ and\ \bibinfo {author} {\bibfnamefont {K.}~\bibnamefont {Semba}},\ }\bibfield  {title} {\bibinfo {title} {Superconducting qubit--oscillator circuit beyond the ultrastrong-coupling regime},\ }\href {https://doi.org/10.1038/nphys3906} {\bibfield  {journal} {\bibinfo  {journal} {Nature Physics}\ }\textbf {\bibinfo {volume} {13}},\ \bibinfo {pages} {44} (\bibinfo {year} {2017}{\natexlab{a}})}\BibitemShut {NoStop}%
\bibitem [{\citenamefont {Yoshihara}\ \emph {et~al.}(2017{\natexlab{b}})\citenamefont {Yoshihara}, \citenamefont {Fuse}, \citenamefont {Ashhab}, \citenamefont {Kakuyanagi}, \citenamefont {Saito},\ and\ \citenamefont {Semba}}]{yoshihara2017pra}%
  \BibitemOpen
  \bibfield  {author} {\bibinfo {author} {\bibfnamefont {F.}~\bibnamefont {Yoshihara}}, \bibinfo {author} {\bibfnamefont {T.}~\bibnamefont {Fuse}}, \bibinfo {author} {\bibfnamefont {S.}~\bibnamefont {Ashhab}}, \bibinfo {author} {\bibfnamefont {K.}~\bibnamefont {Kakuyanagi}}, \bibinfo {author} {\bibfnamefont {S.}~\bibnamefont {Saito}},\ and\ \bibinfo {author} {\bibfnamefont {K.}~\bibnamefont {Semba}},\ }\bibfield  {title} {\bibinfo {title} {Characteristic spectra of circuit quantum electrodynamics systems from the ultrastrong- to the deep-strong-coupling regime},\ }\href {https://doi.org/10.1103/PhysRevA.95.053824} {\bibfield  {journal} {\bibinfo  {journal} {Phys. Rev. A}\ }\textbf {\bibinfo {volume} {95}},\ \bibinfo {pages} {053824} (\bibinfo {year} {2017}{\natexlab{b}})}\BibitemShut {NoStop}%
\bibitem [{\citenamefont {Deng}\ \emph {et~al.}(2015)\citenamefont {Deng}, \citenamefont {Orgiazzi}, \citenamefont {Shen}, \citenamefont {Ashhab},\ and\ \citenamefont {Lupascu}}]{deng2015prl}%
  \BibitemOpen
  \bibfield  {author} {\bibinfo {author} {\bibfnamefont {C.}~\bibnamefont {Deng}}, \bibinfo {author} {\bibfnamefont {J.-L.}\ \bibnamefont {Orgiazzi}}, \bibinfo {author} {\bibfnamefont {F.}~\bibnamefont {Shen}}, \bibinfo {author} {\bibfnamefont {S.}~\bibnamefont {Ashhab}},\ and\ \bibinfo {author} {\bibfnamefont {A.}~\bibnamefont {Lupascu}},\ }\bibfield  {title} {\bibinfo {title} {Observation of floquet states in a strongly driven artificial atom},\ }\href {https://doi.org/10.1103/PhysRevLett.115.133601} {\bibfield  {journal} {\bibinfo  {journal} {Phys. Rev. Lett.}\ }\textbf {\bibinfo {volume} {115}},\ \bibinfo {pages} {133601} (\bibinfo {year} {2015})}\BibitemShut {NoStop}%
\bibitem [{\citenamefont {Hofheinz}\ \emph {et~al.}(2008)\citenamefont {Hofheinz}, \citenamefont {Weig}, \citenamefont {Ansmann}, \citenamefont {Bialczak}, \citenamefont {Lucero}, \citenamefont {Neeley}, \citenamefont {O'Connell}, \citenamefont {Wang}, \citenamefont {Martinis},\ and\ \citenamefont {Cleland}}]{Hofheinz2008}%
  \BibitemOpen
  \bibfield  {author} {\bibinfo {author} {\bibfnamefont {M.}~\bibnamefont {Hofheinz}}, \bibinfo {author} {\bibfnamefont {E.~M.}\ \bibnamefont {Weig}}, \bibinfo {author} {\bibfnamefont {M.}~\bibnamefont {Ansmann}}, \bibinfo {author} {\bibfnamefont {R.~C.}\ \bibnamefont {Bialczak}}, \bibinfo {author} {\bibfnamefont {E.}~\bibnamefont {Lucero}}, \bibinfo {author} {\bibfnamefont {M.}~\bibnamefont {Neeley}}, \bibinfo {author} {\bibfnamefont {A.~D.}\ \bibnamefont {O'Connell}}, \bibinfo {author} {\bibfnamefont {H.}~\bibnamefont {Wang}}, \bibinfo {author} {\bibfnamefont {J.~M.}\ \bibnamefont {Martinis}},\ and\ \bibinfo {author} {\bibfnamefont {A.~N.}\ \bibnamefont {Cleland}},\ }\bibfield  {title} {\bibinfo {title} {Generation of fock states in a superconducting quantum circuit},\ }\href {https://doi.org/10.1038/nature07136} {\bibfield  {journal} {\bibinfo  {journal} {Nature}\ }\textbf {\bibinfo {volume} {454}},\ \bibinfo {pages} {310} (\bibinfo {year} {2008})}\BibitemShut {NoStop}%
\bibitem [{\citenamefont {Kjaergaard}\ \emph {et~al.}(2020)\citenamefont {Kjaergaard}, \citenamefont {Schwartz}, \citenamefont {Braumüller}, \citenamefont {Krantz}, \citenamefont {Wang}, \citenamefont {Gustavsson},\ and\ \citenamefont {Oliver}}]{kjaergaard2019annrev}%
  \BibitemOpen
  \bibfield  {author} {\bibinfo {author} {\bibfnamefont {M.}~\bibnamefont {Kjaergaard}}, \bibinfo {author} {\bibfnamefont {M.~E.}\ \bibnamefont {Schwartz}}, \bibinfo {author} {\bibfnamefont {J.}~\bibnamefont {Braumüller}}, \bibinfo {author} {\bibfnamefont {P.}~\bibnamefont {Krantz}}, \bibinfo {author} {\bibfnamefont {J.~I.-J.}\ \bibnamefont {Wang}}, \bibinfo {author} {\bibfnamefont {S.}~\bibnamefont {Gustavsson}},\ and\ \bibinfo {author} {\bibfnamefont {W.~D.}\ \bibnamefont {Oliver}},\ }\bibfield  {title} {\bibinfo {title} {Superconducting qubits: Current state of play},\ }\href {https://doi.org/https://doi.org/10.1146/annurev-conmatphys-031119-050605} {\bibfield  {journal} {\bibinfo  {journal} {Annual Review of Condensed Matter Physics}\ }\textbf {\bibinfo {volume} {11}},\ \bibinfo {pages} {369} (\bibinfo {year} {2020})}\BibitemShut {NoStop}%
\bibitem [{\citenamefont {Krantz}\ \emph {et~al.}(2019)\citenamefont {Krantz}, \citenamefont {Kjaergaard}, \citenamefont {Yan}, \citenamefont {Orlando}, \citenamefont {Gustavsson},\ and\ \citenamefont {Oliver}}]{Krantz2019appphysrev}%
  \BibitemOpen
  \bibfield  {author} {\bibinfo {author} {\bibfnamefont {P.}~\bibnamefont {Krantz}}, \bibinfo {author} {\bibfnamefont {M.}~\bibnamefont {Kjaergaard}}, \bibinfo {author} {\bibfnamefont {F.}~\bibnamefont {Yan}}, \bibinfo {author} {\bibfnamefont {T.~P.}\ \bibnamefont {Orlando}}, \bibinfo {author} {\bibfnamefont {S.}~\bibnamefont {Gustavsson}},\ and\ \bibinfo {author} {\bibfnamefont {W.~D.}\ \bibnamefont {Oliver}},\ }\bibfield  {title} {\bibinfo {title} {A quantum engineer's guide to superconducting qubits},\ }\href {https://doi.org/10.1063/1.5089550} {\bibfield  {journal} {\bibinfo  {journal} {Applied Physics Reviews}\ }\textbf {\bibinfo {volume} {6}},\ \bibinfo {pages} {021318} (\bibinfo {year} {2019})}\BibitemShut {NoStop}%
\bibitem [{\citenamefont {Leghtas}\ \emph {et~al.}(2013)\citenamefont {Leghtas}, \citenamefont {Kirchmair}, \citenamefont {Vlastakis}, \citenamefont {Schoelkopf}, \citenamefont {Devoret},\ and\ \citenamefont {Mirrahimi}}]{PhysRevLett.111.120501}%
  \BibitemOpen
  \bibfield  {author} {\bibinfo {author} {\bibfnamefont {Z.}~\bibnamefont {Leghtas}}, \bibinfo {author} {\bibfnamefont {G.}~\bibnamefont {Kirchmair}}, \bibinfo {author} {\bibfnamefont {B.}~\bibnamefont {Vlastakis}}, \bibinfo {author} {\bibfnamefont {R.~J.}\ \bibnamefont {Schoelkopf}}, \bibinfo {author} {\bibfnamefont {M.~H.}\ \bibnamefont {Devoret}},\ and\ \bibinfo {author} {\bibfnamefont {M.}~\bibnamefont {Mirrahimi}},\ }\bibfield  {title} {\bibinfo {title} {Hardware-efficient autonomous quantum memory protection},\ }\href {https://doi.org/10.1103/PhysRevLett.111.120501} {\bibfield  {journal} {\bibinfo  {journal} {Phys. Rev. Lett.}\ }\textbf {\bibinfo {volume} {111}},\ \bibinfo {pages} {120501} (\bibinfo {year} {2013})}\BibitemShut {NoStop}%
\bibitem [{\citenamefont {Niemczyk}\ \emph {et~al.}(2010)\citenamefont {Niemczyk}, \citenamefont {Deppe}, \citenamefont {Huebl}, \citenamefont {Menzel}, \citenamefont {Hocke}, \citenamefont {Schwarz}, \citenamefont {Garcia-Ripoll}, \citenamefont {Zueco}, \citenamefont {H{\"u}mmer}, \citenamefont {Solano}, \citenamefont {Marx},\ and\ \citenamefont {Gross}}]{Niemczyk2010}%
  \BibitemOpen
  \bibfield  {author} {\bibinfo {author} {\bibfnamefont {T.}~\bibnamefont {Niemczyk}}, \bibinfo {author} {\bibfnamefont {F.}~\bibnamefont {Deppe}}, \bibinfo {author} {\bibfnamefont {H.}~\bibnamefont {Huebl}}, \bibinfo {author} {\bibfnamefont {E.~P.}\ \bibnamefont {Menzel}}, \bibinfo {author} {\bibfnamefont {F.}~\bibnamefont {Hocke}}, \bibinfo {author} {\bibfnamefont {M.~J.}\ \bibnamefont {Schwarz}}, \bibinfo {author} {\bibfnamefont {J.~J.}\ \bibnamefont {Garcia-Ripoll}}, \bibinfo {author} {\bibfnamefont {D.}~\bibnamefont {Zueco}}, \bibinfo {author} {\bibfnamefont {T.}~\bibnamefont {H{\"u}mmer}}, \bibinfo {author} {\bibfnamefont {E.}~\bibnamefont {Solano}}, \bibinfo {author} {\bibfnamefont {A.}~\bibnamefont {Marx}},\ and\ \bibinfo {author} {\bibfnamefont {R.}~\bibnamefont {Gross}},\ }\bibfield  {title} {\bibinfo {title} {Circuit quantum electrodynamics in the ultrastrong-coupling regime},\ }\href {https://doi.org/10.1038/nphys1730} {\bibfield  {journal} {\bibinfo  {journal} {Nature Physics}\ }\textbf {\bibinfo
  {volume} {6}},\ \bibinfo {pages} {772} (\bibinfo {year} {2010})}\BibitemShut {NoStop}%
\bibitem [{\citenamefont {Fink}\ \emph {et~al.}(2008)\citenamefont {Fink}, \citenamefont {G{\"o}ppl}, \citenamefont {Baur}, \citenamefont {Bianchetti}, \citenamefont {Leek}, \citenamefont {Blais},\ and\ \citenamefont {Wallraff}}]{Fink2008}%
  \BibitemOpen
  \bibfield  {author} {\bibinfo {author} {\bibfnamefont {J.~M.}\ \bibnamefont {Fink}}, \bibinfo {author} {\bibfnamefont {M.}~\bibnamefont {G{\"o}ppl}}, \bibinfo {author} {\bibfnamefont {M.}~\bibnamefont {Baur}}, \bibinfo {author} {\bibfnamefont {R.}~\bibnamefont {Bianchetti}}, \bibinfo {author} {\bibfnamefont {P.~J.}\ \bibnamefont {Leek}}, \bibinfo {author} {\bibfnamefont {A.}~\bibnamefont {Blais}},\ and\ \bibinfo {author} {\bibfnamefont {A.}~\bibnamefont {Wallraff}},\ }\bibfield  {title} {\bibinfo {title} {Climbing the {J}aynes--{C}ummings ladder and observing its nonlinearity in a cavity {QED} system},\ }\href {https://doi.org/10.1038/nature07112} {\bibfield  {journal} {\bibinfo  {journal} {Nature}\ }\textbf {\bibinfo {volume} {454}},\ \bibinfo {pages} {315} (\bibinfo {year} {2008})}\BibitemShut {NoStop}%
\end{thebibliography}%


\begin{thebibliography}{62}%
\makeatletter
\providecommand \@ifxundefined [1]{%
 \@ifx{#1\undefined}
}%
\providecommand \@ifnum [1]{%
 \ifnum #1\expandafter \@firstoftwo
 \else \expandafter \@secondoftwo
 \fi
}%
\providecommand \@ifx [1]{%
 \ifx #1\expandafter \@firstoftwo
 \else \expandafter \@secondoftwo
 \fi
}%
\providecommand \natexlab [1]{#1}%
\providecommand \enquote  [1]{``#1''}%
\providecommand \bibnamefont  [1]{#1}%
\providecommand \bibfnamefont [1]{#1}%
\providecommand \citenamefont [1]{#1}%
\providecommand \href@noop [0]{\@secondoftwo}%
\providecommand \href [0]{\begingroup \@sanitize@url \@href}%
\providecommand \@href[1]{\@@startlink{#1}\@@href}%
\providecommand \@@href[1]{\endgroup#1\@@endlink}%
\providecommand \@sanitize@url [0]{\catcode `\\12\catcode `\$12\catcode `\&12\catcode `\#12\catcode `\^12\catcode `\_12\catcode `\%12\relax}%
\providecommand \@@startlink[1]{}%
\providecommand \@@endlink[0]{}%
\providecommand \url  [0]{\begingroup\@sanitize@url \@url }%
\providecommand \@url [1]{\endgroup\@href {#1}{\urlprefix }}%
\providecommand \urlprefix  [0]{URL }%
\providecommand \Eprint [0]{\href }%
\providecommand \doibase [0]{https://doi.org/}%
\providecommand \selectlanguage [0]{\@gobble}%
\providecommand \bibinfo  [0]{\@secondoftwo}%
\providecommand \bibfield  [0]{\@secondoftwo}%
\providecommand \translation [1]{[#1]}%
\providecommand \BibitemOpen [0]{}%
\providecommand \bibitemStop [0]{}%
\providecommand \bibitemNoStop [0]{.\EOS\space}%
\providecommand \EOS [0]{\spacefactor3000\relax}%
\providecommand \BibitemShut  [1]{\csname bibitem#1\endcsname}%
\let\auto@bib@innerbib\@empty
\bibitem [{\citenamefont {Nielsen}\ and\ \citenamefont {Chuang}(2010)}]{Nielsen_Chuang_2010}%
  \BibitemOpen
  \bibfield  {author} {\bibinfo {author} {\bibfnamefont {M.~A.}\ \bibnamefont {Nielsen}}\ and\ \bibinfo {author} {\bibfnamefont {I.~L.}\ \bibnamefont {Chuang}},\ }\href@noop {} {\emph {\bibinfo {title} {Quantum Computation and Quantum Information: 10th Anniversary Edition}}}\ (\bibinfo  {publisher} {Cambridge University Press},\ \bibinfo {year} {2010})\BibitemShut {NoStop}%
\bibitem [{\citenamefont {Benenti}\ \emph {et~al.}(2018)\citenamefont {Benenti}, \citenamefont {Casati}, \citenamefont {Rossini},\ and\ \citenamefont {Strini}}]{BenentiCasati}%
  \BibitemOpen
  \bibfield  {author} {\bibinfo {author} {\bibfnamefont {G.}~\bibnamefont {Benenti}}, \bibinfo {author} {\bibfnamefont {G.}~\bibnamefont {Casati}}, \bibinfo {author} {\bibfnamefont {D.}~\bibnamefont {Rossini}},\ and\ \bibinfo {author} {\bibfnamefont {G.}~\bibnamefont {Strini}},\ }\href {https://doi.org/10.1142/10909} {\emph {\bibinfo {title} {Principles of Quantum Computation and Information}}},\ \bibinfo {edition} {2nd}\ ed.\ (\bibinfo  {publisher} {World Scientific},\ \bibinfo {address} {Singapore},\ \bibinfo {year} {2018})\BibitemShut {NoStop}%
\bibitem [{\citenamefont {Walls}\ and\ \citenamefont {Milburn}(2008)}]{Walls2008}%
  \BibitemOpen
  \bibfield  {author} {\bibinfo {author} {\bibfnamefont {D.}~\bibnamefont {Walls}}\ and\ \bibinfo {author} {\bibfnamefont {G.~J.}\ \bibnamefont {Milburn}},\ }\bibinfo {title} {Generation and applications of squeezed light},\ in\ \href {https://doi.org/10.1007/978-3-540-28574-8_8} {\emph {\bibinfo {booktitle} {Quantum Optics}}},\ \bibinfo {editor} {edited by\ \bibinfo {editor} {\bibfnamefont {D.}~\bibnamefont {Walls}}\ and\ \bibinfo {editor} {\bibfnamefont {G.~J.}\ \bibnamefont {Milburn}}}\ (\bibinfo  {publisher} {Springer Berlin Heidelberg},\ \bibinfo {address} {Berlin, Heidelberg},\ \bibinfo {year} {2008})\ pp.\ \bibinfo {pages} {143--175}\BibitemShut {NoStop}%
\bibitem [{\citenamefont {Chen}\ \emph {et~al.}(2021{\natexlab{a}})\citenamefont {Chen}, \citenamefont {Qin}, \citenamefont {Wang}, \citenamefont {Miranowicz},\ and\ \citenamefont {Nori}}]{chen2021prl}%
  \BibitemOpen
  \bibfield  {author} {\bibinfo {author} {\bibfnamefont {Y.-H.}\ \bibnamefont {Chen}}, \bibinfo {author} {\bibfnamefont {W.}~\bibnamefont {Qin}}, \bibinfo {author} {\bibfnamefont {X.}~\bibnamefont {Wang}}, \bibinfo {author} {\bibfnamefont {A.}~\bibnamefont {Miranowicz}},\ and\ \bibinfo {author} {\bibfnamefont {F.}~\bibnamefont {Nori}},\ }\bibfield  {title} {\bibinfo {title} {Shortcuts to adiabaticity for the quantum rabi model: Efficient generation of giant entangled cat states via parametric amplification},\ }\href {https://doi.org/10.1103/PhysRevLett.126.023602} {\bibfield  {journal} {\bibinfo  {journal} {Phys. Rev. Lett.}\ }\textbf {\bibinfo {volume} {126}},\ \bibinfo {pages} {023602} (\bibinfo {year} {2021}{\natexlab{a}})}\BibitemShut {NoStop}%
\bibitem [{\citenamefont {Aasi}\ \emph {et~al.}(2013)\citenamefont {Aasi} \emph {et~al.}}]{Aasi2013}%
  \BibitemOpen
  \bibfield  {author} {\bibinfo {author} {\bibfnamefont {J.}~\bibnamefont {Aasi}} \emph {et~al.},\ }\bibfield  {title} {\bibinfo {title} {Enhanced sensitivity of the ligo gravitational wave detector by using squeezed states of light},\ }\href {https://doi.org/10.1038/nphoton.2013.177} {\bibfield  {journal} {\bibinfo  {journal} {Nature Photonics}\ }\textbf {\bibinfo {volume} {7}},\ \bibinfo {pages} {613} (\bibinfo {year} {2013})}\BibitemShut {NoStop}%
\bibitem [{\citenamefont {Descamps}\ \emph {et~al.}(2026)\citenamefont {Descamps}, \citenamefont {Saharyan}, \citenamefont {Chivet}, \citenamefont {Keller},\ and\ \citenamefont {Milman}}]{Descamps26}%
  \BibitemOpen
  \bibfield  {author} {\bibinfo {author} {\bibfnamefont {E.}~\bibnamefont {Descamps}}, \bibinfo {author} {\bibfnamefont {A.}~\bibnamefont {Saharyan}}, \bibinfo {author} {\bibfnamefont {A.}~\bibnamefont {Chivet}}, \bibinfo {author} {\bibfnamefont {A.}~\bibnamefont {Keller}},\ and\ \bibinfo {author} {\bibfnamefont {P.}~\bibnamefont {Milman}},\ }\bibfield  {title} {\bibinfo {title} {Unified framework for bosonic quantum information encoding, resources, and universality from superselection rules},\ }\href {https://doi.org/10.1364/OPTICAQ.581218} {\bibfield  {journal} {\bibinfo  {journal} {Optica Quantum}\ }\textbf {\bibinfo {volume} {4}},\ \bibinfo {pages} {148} (\bibinfo {year} {2026})}\BibitemShut {NoStop}%
\bibitem [{\citenamefont {Kurman}\ \emph {et~al.}(2026)\citenamefont {Kurman}, \citenamefont {Hymas}, \citenamefont {Fedorov}, \citenamefont {Munro},\ and\ \citenamefont {Quach}}]{kurman2026prx}%
  \BibitemOpen
  \bibfield  {author} {\bibinfo {author} {\bibfnamefont {Y.}~\bibnamefont {Kurman}}, \bibinfo {author} {\bibfnamefont {K.}~\bibnamefont {Hymas}}, \bibinfo {author} {\bibfnamefont {A.}~\bibnamefont {Fedorov}}, \bibinfo {author} {\bibfnamefont {W.~J.}\ \bibnamefont {Munro}},\ and\ \bibinfo {author} {\bibfnamefont {J.}~\bibnamefont {Quach}},\ }\bibfield  {title} {\bibinfo {title} {Powering quantum computation with quantum batteries},\ }\href {https://doi.org/10.1103/l39v-jwwz} {\bibfield  {journal} {\bibinfo  {journal} {Phys. Rev. X}\ }\textbf {\bibinfo {volume} {16}},\ \bibinfo {pages} {011016} (\bibinfo {year} {2026})}\BibitemShut {NoStop}%
\bibitem [{\citenamefont {Michael}\ \emph {et~al.}(2016)\citenamefont {Michael}, \citenamefont {Silveri}, \citenamefont {Brierley}, \citenamefont {Albert}, \citenamefont {Salmilehto}, \citenamefont {Jiang},\ and\ \citenamefont {Girvin}}]{michael2016prx}%
  \BibitemOpen
  \bibfield  {author} {\bibinfo {author} {\bibfnamefont {M.~H.}\ \bibnamefont {Michael}}, \bibinfo {author} {\bibfnamefont {M.}~\bibnamefont {Silveri}}, \bibinfo {author} {\bibfnamefont {R.~T.}\ \bibnamefont {Brierley}}, \bibinfo {author} {\bibfnamefont {V.~V.}\ \bibnamefont {Albert}}, \bibinfo {author} {\bibfnamefont {J.}~\bibnamefont {Salmilehto}}, \bibinfo {author} {\bibfnamefont {L.}~\bibnamefont {Jiang}},\ and\ \bibinfo {author} {\bibfnamefont {S.~M.}\ \bibnamefont {Girvin}},\ }\bibfield  {title} {\bibinfo {title} {New class of quantum error-correcting codes for a bosonic mode},\ }\href {https://doi.org/10.1103/PhysRevX.6.031006} {\bibfield  {journal} {\bibinfo  {journal} {Phys. Rev. X}\ }\textbf {\bibinfo {volume} {6}},\ \bibinfo {pages} {031006} (\bibinfo {year} {2016})}\BibitemShut {NoStop}%
\bibitem [{\citenamefont {Chen}\ \emph {et~al.}(2021{\natexlab{b}})\citenamefont {Chen}, \citenamefont {Qin}, \citenamefont {Stassi}, \citenamefont {Wang},\ and\ \citenamefont {Nori}}]{chen2021prr}%
  \BibitemOpen
  \bibfield  {author} {\bibinfo {author} {\bibfnamefont {Y.-H.}\ \bibnamefont {Chen}}, \bibinfo {author} {\bibfnamefont {W.}~\bibnamefont {Qin}}, \bibinfo {author} {\bibfnamefont {R.}~\bibnamefont {Stassi}}, \bibinfo {author} {\bibfnamefont {X.}~\bibnamefont {Wang}},\ and\ \bibinfo {author} {\bibfnamefont {F.}~\bibnamefont {Nori}},\ }\bibfield  {title} {\bibinfo {title} {Fast binomial-code holonomic quantum computation with ultrastrong light-matter coupling},\ }\href {https://doi.org/10.1103/PhysRevResearch.3.033275} {\bibfield  {journal} {\bibinfo  {journal} {Phys. Rev. Res.}\ }\textbf {\bibinfo {volume} {3}},\ \bibinfo {pages} {033275} (\bibinfo {year} {2021}{\natexlab{b}})}\BibitemShut {NoStop}%
\bibitem [{\citenamefont {Leghtas}\ \emph {et~al.}(2013)\citenamefont {Leghtas}, \citenamefont {Kirchmair}, \citenamefont {Vlastakis}, \citenamefont {Schoelkopf}, \citenamefont {Devoret},\ and\ \citenamefont {Mirrahimi}}]{PhysRevLett.111.120501}%
  \BibitemOpen
  \bibfield  {author} {\bibinfo {author} {\bibfnamefont {Z.}~\bibnamefont {Leghtas}}, \bibinfo {author} {\bibfnamefont {G.}~\bibnamefont {Kirchmair}}, \bibinfo {author} {\bibfnamefont {B.}~\bibnamefont {Vlastakis}}, \bibinfo {author} {\bibfnamefont {R.~J.}\ \bibnamefont {Schoelkopf}}, \bibinfo {author} {\bibfnamefont {M.~H.}\ \bibnamefont {Devoret}},\ and\ \bibinfo {author} {\bibfnamefont {M.}~\bibnamefont {Mirrahimi}},\ }\bibfield  {title} {\bibinfo {title} {Hardware-efficient autonomous quantum memory protection},\ }\href {https://doi.org/10.1103/PhysRevLett.111.120501} {\bibfield  {journal} {\bibinfo  {journal} {Phys. Rev. Lett.}\ }\textbf {\bibinfo {volume} {111}},\ \bibinfo {pages} {120501} (\bibinfo {year} {2013})}\BibitemShut {NoStop}%
\bibitem [{\citenamefont {Ofek}\ \emph {et~al.}(2016)\citenamefont {Ofek}, \citenamefont {Petrenko}, \citenamefont {Heeres}, \citenamefont {Reinhold}, \citenamefont {Leghtas}, \citenamefont {Vlastakis}, \citenamefont {Liu}, \citenamefont {Frunzio}, \citenamefont {Girvin}, \citenamefont {Jiang}, \citenamefont {Mirrahimi}, \citenamefont {Devoret},\ and\ \citenamefont {Schoelkopf}}]{Ofek2016nature}%
  \BibitemOpen
  \bibfield  {author} {\bibinfo {author} {\bibfnamefont {N.}~\bibnamefont {Ofek}}, \bibinfo {author} {\bibfnamefont {A.}~\bibnamefont {Petrenko}}, \bibinfo {author} {\bibfnamefont {R.}~\bibnamefont {Heeres}}, \bibinfo {author} {\bibfnamefont {P.}~\bibnamefont {Reinhold}}, \bibinfo {author} {\bibfnamefont {Z.}~\bibnamefont {Leghtas}}, \bibinfo {author} {\bibfnamefont {B.}~\bibnamefont {Vlastakis}}, \bibinfo {author} {\bibfnamefont {Y.}~\bibnamefont {Liu}}, \bibinfo {author} {\bibfnamefont {L.}~\bibnamefont {Frunzio}}, \bibinfo {author} {\bibfnamefont {S.~M.}\ \bibnamefont {Girvin}}, \bibinfo {author} {\bibfnamefont {L.}~\bibnamefont {Jiang}}, \bibinfo {author} {\bibfnamefont {M.}~\bibnamefont {Mirrahimi}}, \bibinfo {author} {\bibfnamefont {M.~H.}\ \bibnamefont {Devoret}},\ and\ \bibinfo {author} {\bibfnamefont {R.~J.}\ \bibnamefont {Schoelkopf}},\ }\bibfield  {title} {\bibinfo {title} {Extending the lifetime of a quantum bit with error correction in superconducting circuits},\ }\href
  {https://doi.org/10.1038/nature18949} {\bibfield  {journal} {\bibinfo  {journal} {Nature}\ }\textbf {\bibinfo {volume} {536}},\ \bibinfo {pages} {441} (\bibinfo {year} {2016})}\BibitemShut {NoStop}%
\bibitem [{\citenamefont {Beaudoin}\ \emph {et~al.}(2011)\citenamefont {Beaudoin}, \citenamefont {Gambetta},\ and\ \citenamefont {Blais}}]{Beaudoin2011}%
  \BibitemOpen
  \bibfield  {author} {\bibinfo {author} {\bibfnamefont {F.}~\bibnamefont {Beaudoin}}, \bibinfo {author} {\bibfnamefont {J.~M.}\ \bibnamefont {Gambetta}},\ and\ \bibinfo {author} {\bibfnamefont {A.}~\bibnamefont {Blais}},\ }\bibfield  {title} {\bibinfo {title} {Dissipation and ultrastrong coupling in circuit {QED}},\ }\href {https://doi.org/10.1103/PhysRevA.84.043832} {\bibfield  {journal} {\bibinfo  {journal} {Phys. Rev. A}\ }\textbf {\bibinfo {volume} {84}},\ \bibinfo {pages} {043832} (\bibinfo {year} {2011})}\BibitemShut {NoStop}%
\bibitem [{\citenamefont {Frisk~Kockum}\ \emph {et~al.}(2019)\citenamefont {Frisk~Kockum}, \citenamefont {Miranowicz}, \citenamefont {De~Liberato}, \citenamefont {Savasta},\ and\ \citenamefont {Nori}}]{Kockum2019}%
  \BibitemOpen
  \bibfield  {author} {\bibinfo {author} {\bibfnamefont {A.}~\bibnamefont {Frisk~Kockum}}, \bibinfo {author} {\bibfnamefont {A.}~\bibnamefont {Miranowicz}}, \bibinfo {author} {\bibfnamefont {S.}~\bibnamefont {De~Liberato}}, \bibinfo {author} {\bibfnamefont {S.}~\bibnamefont {Savasta}},\ and\ \bibinfo {author} {\bibfnamefont {F.}~\bibnamefont {Nori}},\ }\bibfield  {title} {\bibinfo {title} {Ultrastrong coupling between light and matter},\ }\href {https://doi.org/10.1038/s42254-018-0006-2} {\bibfield  {journal} {\bibinfo  {journal} {Nature Reviews Physics}\ }\textbf {\bibinfo {volume} {1}},\ \bibinfo {pages} {19} (\bibinfo {year} {2019})}\BibitemShut {NoStop}%
\bibitem [{\citenamefont {Forn-D\'{\i}az}\ \emph {et~al.}(2019)\citenamefont {Forn-D\'{\i}az}, \citenamefont {Lamata}, \citenamefont {Rico}, \citenamefont {Kono},\ and\ \citenamefont {Solano}}]{forndiaz2019}%
  \BibitemOpen
  \bibfield  {author} {\bibinfo {author} {\bibfnamefont {P.}~\bibnamefont {Forn-D\'{\i}az}}, \bibinfo {author} {\bibfnamefont {L.}~\bibnamefont {Lamata}}, \bibinfo {author} {\bibfnamefont {E.}~\bibnamefont {Rico}}, \bibinfo {author} {\bibfnamefont {J.}~\bibnamefont {Kono}},\ and\ \bibinfo {author} {\bibfnamefont {E.}~\bibnamefont {Solano}},\ }\bibfield  {title} {\bibinfo {title} {Ultrastrong coupling regimes of light-matter interaction},\ }\href {https://doi.org/10.1103/RevModPhys.91.025005} {\bibfield  {journal} {\bibinfo  {journal} {Rev. Mod. Phys.}\ }\textbf {\bibinfo {volume} {91}},\ \bibinfo {pages} {025005} (\bibinfo {year} {2019})}\BibitemShut {NoStop}%
\bibitem [{\citenamefont {Moore}(1970)}]{Moore1970JMathPhys}%
  \BibitemOpen
  \bibfield  {author} {\bibinfo {author} {\bibfnamefont {G.~T.}\ \bibnamefont {Moore}},\ }\bibfield  {title} {\bibinfo {title} {Quantum theory of the electromagnetic field in a variable‐length one‐dimensional cavity},\ }\href {https://doi.org/10.1063/1.1665432} {\bibfield  {journal} {\bibinfo  {journal} {Journal of Mathematical Physics}\ }\textbf {\bibinfo {volume} {11}},\ \bibinfo {pages} {2679} (\bibinfo {year} {1970})}\BibitemShut {NoStop}%
\bibitem [{\citenamefont {Dodonov}(2010)}]{Dodonov_2010}%
  \BibitemOpen
  \bibfield  {author} {\bibinfo {author} {\bibfnamefont {V.~V.}\ \bibnamefont {Dodonov}},\ }\bibfield  {title} {\bibinfo {title} {Current status of the dynamical casimir effect},\ }\href {https://doi.org/10.1088/0031-8949/82/03/038105} {\bibfield  {journal} {\bibinfo  {journal} {Physica Scripta}\ }\textbf {\bibinfo {volume} {82}},\ \bibinfo {pages} {038105} (\bibinfo {year} {2010})}\BibitemShut {NoStop}%
\bibitem [{\citenamefont {Wilson}\ \emph {et~al.}(2011)\citenamefont {Wilson}, \citenamefont {Johansson}, \citenamefont {Pourkabirian}, \citenamefont {Simoen}, \citenamefont {Johansson}, \citenamefont {Duty}, \citenamefont {Nori},\ and\ \citenamefont {Delsing}}]{Wilson2011}%
  \BibitemOpen
  \bibfield  {author} {\bibinfo {author} {\bibfnamefont {C.~M.}\ \bibnamefont {Wilson}}, \bibinfo {author} {\bibfnamefont {G.}~\bibnamefont {Johansson}}, \bibinfo {author} {\bibfnamefont {A.}~\bibnamefont {Pourkabirian}}, \bibinfo {author} {\bibfnamefont {M.}~\bibnamefont {Simoen}}, \bibinfo {author} {\bibfnamefont {J.~R.}\ \bibnamefont {Johansson}}, \bibinfo {author} {\bibfnamefont {T.}~\bibnamefont {Duty}}, \bibinfo {author} {\bibfnamefont {F.}~\bibnamefont {Nori}},\ and\ \bibinfo {author} {\bibfnamefont {P.}~\bibnamefont {Delsing}},\ }\bibfield  {title} {\bibinfo {title} {Observation of the dynamical casimir effect in a superconducting circuit},\ }\href {https://doi.org/10.1038/nature10561} {\bibfield  {journal} {\bibinfo  {journal} {Nature}\ }\textbf {\bibinfo {volume} {479}},\ \bibinfo {pages} {376} (\bibinfo {year} {2011})}\BibitemShut {NoStop}%
\bibitem [{\citenamefont {Nation}\ \emph {et~al.}(2012)\citenamefont {Nation}, \citenamefont {Johansson}, \citenamefont {Blencowe},\ and\ \citenamefont {Nori}}]{RevModPhys.84.1}%
  \BibitemOpen
  \bibfield  {author} {\bibinfo {author} {\bibfnamefont {P.~D.}\ \bibnamefont {Nation}}, \bibinfo {author} {\bibfnamefont {J.~R.}\ \bibnamefont {Johansson}}, \bibinfo {author} {\bibfnamefont {M.~P.}\ \bibnamefont {Blencowe}},\ and\ \bibinfo {author} {\bibfnamefont {F.}~\bibnamefont {Nori}},\ }\bibfield  {title} {\bibinfo {title} {Colloquium: Stimulating uncertainty: Amplifying the quantum vacuum with superconducting circuits},\ }\href {https://doi.org/10.1103/RevModPhys.84.1} {\bibfield  {journal} {\bibinfo  {journal} {Rev. Mod. Phys.}\ }\textbf {\bibinfo {volume} {84}},\ \bibinfo {pages} {1} (\bibinfo {year} {2012})}\BibitemShut {NoStop}%
\bibitem [{\citenamefont {Hoeb}\ \emph {et~al.}(2017)\citenamefont {Hoeb}, \citenamefont {Angaroni}, \citenamefont {Zoller}, \citenamefont {Calarco}, \citenamefont {Strini}, \citenamefont {Montangero},\ and\ \citenamefont {Benenti}}]{Hoeb2017}%
  \BibitemOpen
  \bibfield  {author} {\bibinfo {author} {\bibfnamefont {F.}~\bibnamefont {Hoeb}}, \bibinfo {author} {\bibfnamefont {F.}~\bibnamefont {Angaroni}}, \bibinfo {author} {\bibfnamefont {J.}~\bibnamefont {Zoller}}, \bibinfo {author} {\bibfnamefont {T.}~\bibnamefont {Calarco}}, \bibinfo {author} {\bibfnamefont {G.}~\bibnamefont {Strini}}, \bibinfo {author} {\bibfnamefont {S.}~\bibnamefont {Montangero}},\ and\ \bibinfo {author} {\bibfnamefont {G.}~\bibnamefont {Benenti}},\ }\bibfield  {title} {\bibinfo {title} {Amplification of the parametric dynamical casimir effect via optimal control},\ }\href {https://doi.org/10.1103/PhysRevA.96.033851} {\bibfield  {journal} {\bibinfo  {journal} {Phys. Rev. A}\ }\textbf {\bibinfo {volume} {96}},\ \bibinfo {pages} {033851} (\bibinfo {year} {2017})}\BibitemShut {NoStop}%
\bibitem [{\citenamefont {Benenti}\ \emph {et~al.}(2014)\citenamefont {Benenti}, \citenamefont {D'Arrigo}, \citenamefont {Siccardi},\ and\ \citenamefont {Strini}}]{PhysRevA.90.052313}%
  \BibitemOpen
  \bibfield  {author} {\bibinfo {author} {\bibfnamefont {G.}~\bibnamefont {Benenti}}, \bibinfo {author} {\bibfnamefont {A.}~\bibnamefont {D'Arrigo}}, \bibinfo {author} {\bibfnamefont {S.}~\bibnamefont {Siccardi}},\ and\ \bibinfo {author} {\bibfnamefont {G.}~\bibnamefont {Strini}},\ }\bibfield  {title} {\bibinfo {title} {Dynamical casimir effect in quantum-information processing},\ }\href {https://doi.org/10.1103/PhysRevA.90.052313} {\bibfield  {journal} {\bibinfo  {journal} {Phys. Rev. A}\ }\textbf {\bibinfo {volume} {90}},\ \bibinfo {pages} {052313} (\bibinfo {year} {2014})}\BibitemShut {NoStop}%
\bibitem [{\citenamefont {Felicetti}\ \emph {et~al.}(2014)\citenamefont {Felicetti}, \citenamefont {Sanz}, \citenamefont {Lamata}, \citenamefont {Romero}, \citenamefont {Johansson}, \citenamefont {Delsing},\ and\ \citenamefont {Solano}}]{felicetti2014prl}%
  \BibitemOpen
  \bibfield  {author} {\bibinfo {author} {\bibfnamefont {S.}~\bibnamefont {Felicetti}}, \bibinfo {author} {\bibfnamefont {M.}~\bibnamefont {Sanz}}, \bibinfo {author} {\bibfnamefont {L.}~\bibnamefont {Lamata}}, \bibinfo {author} {\bibfnamefont {G.}~\bibnamefont {Romero}}, \bibinfo {author} {\bibfnamefont {G.}~\bibnamefont {Johansson}}, \bibinfo {author} {\bibfnamefont {P.}~\bibnamefont {Delsing}},\ and\ \bibinfo {author} {\bibfnamefont {E.}~\bibnamefont {Solano}},\ }\bibfield  {title} {\bibinfo {title} {Dynamical casimir effect entangles artificial atoms},\ }\href {https://doi.org/10.1103/PhysRevLett.113.093602} {\bibfield  {journal} {\bibinfo  {journal} {Phys. Rev. Lett.}\ }\textbf {\bibinfo {volume} {113}},\ \bibinfo {pages} {093602} (\bibinfo {year} {2014})}\BibitemShut {NoStop}%
\bibitem [{\citenamefont {Koch}\ \emph {et~al.}(2022)\citenamefont {Koch}, \citenamefont {Boscain}, \citenamefont {Calarco}, \citenamefont {Dirr}, \citenamefont {Filipp}, \citenamefont {Glaser}, \citenamefont {Kosloff}, \citenamefont {Montangero}, \citenamefont {Schulte-Herbr{\"u}ggen}, \citenamefont {Sugny},\ and\ \citenamefont {Wilhelm}}]{Koch2022}%
  \BibitemOpen
  \bibfield  {author} {\bibinfo {author} {\bibfnamefont {C.~P.}\ \bibnamefont {Koch}}, \bibinfo {author} {\bibfnamefont {U.}~\bibnamefont {Boscain}}, \bibinfo {author} {\bibfnamefont {T.}~\bibnamefont {Calarco}}, \bibinfo {author} {\bibfnamefont {G.}~\bibnamefont {Dirr}}, \bibinfo {author} {\bibfnamefont {S.}~\bibnamefont {Filipp}}, \bibinfo {author} {\bibfnamefont {S.~J.}\ \bibnamefont {Glaser}}, \bibinfo {author} {\bibfnamefont {R.}~\bibnamefont {Kosloff}}, \bibinfo {author} {\bibfnamefont {S.}~\bibnamefont {Montangero}}, \bibinfo {author} {\bibfnamefont {T.}~\bibnamefont {Schulte-Herbr{\"u}ggen}}, \bibinfo {author} {\bibfnamefont {D.}~\bibnamefont {Sugny}},\ and\ \bibinfo {author} {\bibfnamefont {F.~K.}\ \bibnamefont {Wilhelm}},\ }\bibfield  {title} {\bibinfo {title} {Quantum optimal control in quantum technologies. {S}trategic report on current status, visions and goals for research in {E}urope},\ }\href {https://doi.org/10.1140/epjqt/s40507-022-00138-x} {\bibfield  {journal} {\bibinfo  {journal} {EPJ
  Quantum Technology}\ }\textbf {\bibinfo {volume} {9}},\ \bibinfo {pages} {19} (\bibinfo {year} {2022})}\BibitemShut {NoStop}%
\bibitem [{\citenamefont {Caneva}\ \emph {et~al.}(2011)\citenamefont {Caneva}, \citenamefont {Calarco},\ and\ \citenamefont {Montangero}}]{PhysRevA.84.022326}%
  \BibitemOpen
  \bibfield  {author} {\bibinfo {author} {\bibfnamefont {T.}~\bibnamefont {Caneva}}, \bibinfo {author} {\bibfnamefont {T.}~\bibnamefont {Calarco}},\ and\ \bibinfo {author} {\bibfnamefont {S.}~\bibnamefont {Montangero}},\ }\bibfield  {title} {\bibinfo {title} {Chopped random-basis quantum optimization},\ }\href {https://doi.org/10.1103/PhysRevA.84.022326} {\bibfield  {journal} {\bibinfo  {journal} {Phys. Rev. A}\ }\textbf {\bibinfo {volume} {84}},\ \bibinfo {pages} {022326} (\bibinfo {year} {2011})}\BibitemShut {NoStop}%
\bibitem [{\citenamefont {Müller}\ \emph {et~al.}(2022)\citenamefont {Müller}, \citenamefont {Said}, \citenamefont {Jelezko}, \citenamefont {Calarco},\ and\ \citenamefont {Montangero}}]{Muller2022}%
  \BibitemOpen
  \bibfield  {author} {\bibinfo {author} {\bibfnamefont {M.~M.}\ \bibnamefont {Müller}}, \bibinfo {author} {\bibfnamefont {R.~S.}\ \bibnamefont {Said}}, \bibinfo {author} {\bibfnamefont {F.}~\bibnamefont {Jelezko}}, \bibinfo {author} {\bibfnamefont {T.}~\bibnamefont {Calarco}},\ and\ \bibinfo {author} {\bibfnamefont {S.}~\bibnamefont {Montangero}},\ }\bibfield  {title} {\bibinfo {title} {One decade of quantum optimal control in the chopped random basis},\ }\href {https://doi.org/10.1088/1361-6633/ac723c} {\bibfield  {journal} {\bibinfo  {journal} {Reports on Progress in Physics}\ }\textbf {\bibinfo {volume} {85}},\ \bibinfo {pages} {076001} (\bibinfo {year} {2022})}\BibitemShut {NoStop}%
\bibitem [{\citenamefont {Khaneja}\ \emph {et~al.}(2005)\citenamefont {Khaneja}, \citenamefont {Reiss}, \citenamefont {Kehlet}, \citenamefont {Schulte-Herbrüggen},\ and\ \citenamefont {Glaser}}]{KHANEJA2005296}%
  \BibitemOpen
  \bibfield  {author} {\bibinfo {author} {\bibfnamefont {N.}~\bibnamefont {Khaneja}}, \bibinfo {author} {\bibfnamefont {T.}~\bibnamefont {Reiss}}, \bibinfo {author} {\bibfnamefont {C.}~\bibnamefont {Kehlet}}, \bibinfo {author} {\bibfnamefont {T.}~\bibnamefont {Schulte-Herbrüggen}},\ and\ \bibinfo {author} {\bibfnamefont {S.~J.}\ \bibnamefont {Glaser}},\ }\bibfield  {title} {\bibinfo {title} {Optimal control of coupled spin dynamics: design of nmr pulse sequences by gradient ascent algorithms},\ }\href {https://doi.org/https://doi.org/10.1016/j.jmr.2004.11.004} {\bibfield  {journal} {\bibinfo  {journal} {Journal of Magnetic Resonance}\ }\textbf {\bibinfo {volume} {172}},\ \bibinfo {pages} {296} (\bibinfo {year} {2005})}\BibitemShut {NoStop}%
\bibitem [{\citenamefont {Haroche}\ and\ \citenamefont {Raimond}(2006)}]{Harochebook}%
  \BibitemOpen
  \bibfield  {author} {\bibinfo {author} {\bibfnamefont {S.}~\bibnamefont {Haroche}}\ and\ \bibinfo {author} {\bibfnamefont {J.-M.}\ \bibnamefont {Raimond}},\ }\href {https://doi.org/10.1093/acprof:oso/9780198509141.001.0001} {\emph {\bibinfo {title} {Exploring the Quantum: Atoms, Cavities, and Photons}}}\ (\bibinfo  {publisher} {Oxford University Press},\ \bibinfo {year} {2006})\BibitemShut {NoStop}%
\bibitem [{\citenamefont {Filipowicz}\ \emph {et~al.}(1986)\citenamefont {Filipowicz}, \citenamefont {Javanainen},\ and\ \citenamefont {Meystre}}]{Filipowicz:86}%
  \BibitemOpen
  \bibfield  {author} {\bibinfo {author} {\bibfnamefont {P.}~\bibnamefont {Filipowicz}}, \bibinfo {author} {\bibfnamefont {J.}~\bibnamefont {Javanainen}},\ and\ \bibinfo {author} {\bibfnamefont {P.}~\bibnamefont {Meystre}},\ }\bibfield  {title} {\bibinfo {title} {Quantum and semiclassical steady states of a kicked cavity mode},\ }\href {https://doi.org/10.1364/JOSAB.3.000906} {\bibfield  {journal} {\bibinfo  {journal} {J. Opt. Soc. Am. B}\ }\textbf {\bibinfo {volume} {3}},\ \bibinfo {pages} {906} (\bibinfo {year} {1986})}\BibitemShut {NoStop}%
\bibitem [{\citenamefont {Braak}(2011)}]{braak2011prl}%
  \BibitemOpen
  \bibfield  {author} {\bibinfo {author} {\bibfnamefont {D.}~\bibnamefont {Braak}},\ }\bibfield  {title} {\bibinfo {title} {Integrability of the {R}abi model},\ }\href {https://doi.org/10.1103/PhysRevLett.107.100401} {\bibfield  {journal} {\bibinfo  {journal} {Phys. Rev. Lett.}\ }\textbf {\bibinfo {volume} {107}},\ \bibinfo {pages} {100401} (\bibinfo {year} {2011})}\BibitemShut {NoStop}%
\bibitem [{\citenamefont {Xie}\ \emph {et~al.}(2017)\citenamefont {Xie}, \citenamefont {Zhong}, \citenamefont {Batchelor},\ and\ \citenamefont {Lee}}]{xie2017jphysa}%
  \BibitemOpen
  \bibfield  {author} {\bibinfo {author} {\bibfnamefont {Q.}~\bibnamefont {Xie}}, \bibinfo {author} {\bibfnamefont {H.}~\bibnamefont {Zhong}}, \bibinfo {author} {\bibfnamefont {M.~T.}\ \bibnamefont {Batchelor}},\ and\ \bibinfo {author} {\bibfnamefont {C.}~\bibnamefont {Lee}},\ }\bibfield  {title} {\bibinfo {title} {The quantum rabi model: solution and dynamics},\ }\href {https://doi.org/10.1088/1751-8121/aa5a65} {\bibfield  {journal} {\bibinfo  {journal} {Journal of Physics A: Mathematical and Theoretical}\ }\textbf {\bibinfo {volume} {50}},\ \bibinfo {pages} {113001} (\bibinfo {year} {2017})}\BibitemShut {NoStop}%
\bibitem [{\citenamefont {Dodonov}(2020)}]{dodonov2020physics}%
  \BibitemOpen
  \bibfield  {author} {\bibinfo {author} {\bibfnamefont {V.}~\bibnamefont {Dodonov}},\ }\bibfield  {title} {\bibinfo {title} {Fifty years of the dynamical casimir effect},\ }\href {https://doi.org/10.3390/physics2010007} {\bibfield  {journal} {\bibinfo  {journal} {Physics}\ }\textbf {\bibinfo {volume} {2}},\ \bibinfo {pages} {67} (\bibinfo {year} {2020})}\BibitemShut {NoStop}%
\bibitem [{Note1()}]{Note1}%
  \BibitemOpen
  \bibinfo {note} {In this work, we consider the \protect \textit {parametric} DCE, where photons are generated through the parametric amplification of vacuum fluctuations without moving or changing the boundaries \cite {Dodonov_2010}}\BibitemShut {NoStop}%
\bibitem [{\citenamefont {Braak}(2019)}]{braak2019symmetry}%
  \BibitemOpen
  \bibfield  {author} {\bibinfo {author} {\bibfnamefont {D.}~\bibnamefont {Braak}},\ }\bibfield  {title} {\bibinfo {title} {Symmetries in the quantum rabi model},\ }\href {https://doi.org/10.3390/sym11101259} {\bibfield  {journal} {\bibinfo  {journal} {Symmetry}\ }\textbf {\bibinfo {volume} {11}},\ \bibinfo {pages} {1259} (\bibinfo {year} {2019})}\BibitemShut {NoStop}%
\bibitem [{Note2()}]{Note2}%
  \BibitemOpen
  \bibinfo {note} {Further details on the specific fidelity definitions used---which depend on the adopted optimal control technique---are provided in the Supplemental Material (SM).}\BibitemShut {Stop}%
\bibitem [{\citenamefont {Zheng}\ \emph {et~al.}(2016)\citenamefont {Zheng}, \citenamefont {Campbell}, \citenamefont {De~Chiara},\ and\ \citenamefont {Poletti}}]{PhysRevA.94.042132}%
  \BibitemOpen
  \bibfield  {author} {\bibinfo {author} {\bibfnamefont {Y.}~\bibnamefont {Zheng}}, \bibinfo {author} {\bibfnamefont {S.}~\bibnamefont {Campbell}}, \bibinfo {author} {\bibfnamefont {G.}~\bibnamefont {De~Chiara}},\ and\ \bibinfo {author} {\bibfnamefont {D.}~\bibnamefont {Poletti}},\ }\bibfield  {title} {\bibinfo {title} {Cost of counterdiabatic driving and work output},\ }\href {https://doi.org/10.1103/PhysRevA.94.042132} {\bibfield  {journal} {\bibinfo  {journal} {Phys. Rev. A}\ }\textbf {\bibinfo {volume} {94}},\ \bibinfo {pages} {042132} (\bibinfo {year} {2016})}\BibitemShut {NoStop}%
\bibitem [{\citenamefont {Campbell}\ and\ \citenamefont {Deffner}(2017)}]{campbell2017prl}%
  \BibitemOpen
  \bibfield  {author} {\bibinfo {author} {\bibfnamefont {S.}~\bibnamefont {Campbell}}\ and\ \bibinfo {author} {\bibfnamefont {S.}~\bibnamefont {Deffner}},\ }\bibfield  {title} {\bibinfo {title} {Trade-off between speed and cost in shortcuts to adiabaticity},\ }\href {https://doi.org/10.1103/PhysRevLett.118.100601} {\bibfield  {journal} {\bibinfo  {journal} {Phys. Rev. Lett.}\ }\textbf {\bibinfo {volume} {118}},\ \bibinfo {pages} {100601} (\bibinfo {year} {2017})}\BibitemShut {NoStop}%
\bibitem [{Note3()}]{Note3}%
  \BibitemOpen
  \bibinfo {note} {See \cite {tumbiolo2026prl} for the use of this metric beyond the counterdiabatic-driving regime.}\BibitemShut {Stop}%
\bibitem [{Note4()}]{Note4}%
  \BibitemOpen
  \bibinfo {note} {The timescale $\tau _s$ corresponds to the time required to perform half a vacuum Rabi oscillation in the resonant ($\omega _q=\omega _c$) Jaynes-Cummings limit [i.e., neglecting counter-rotating terms in the Rabi Hamiltonian (\ref {eq:H_Rabi})].}\BibitemShut {Stop}%
\bibitem [{Note5()}]{Note5}%
  \BibitemOpen
  \bibinfo {note} {Depending on the specific experimental constraints, the fidelity $F \approx 0.995$ can be further enhanced up to $\approx 0.999$ by permitting larger control amplitudes (see SM).}\BibitemShut {Stop}%
\bibitem [{\citenamefont {Breuer}\ and\ \citenamefont {Petruccione}(2007)}]{breuer2002theory}%
  \BibitemOpen
  \bibfield  {author} {\bibinfo {author} {\bibfnamefont {H.-P.}\ \bibnamefont {Breuer}}\ and\ \bibinfo {author} {\bibfnamefont {F.}~\bibnamefont {Petruccione}},\ }\href {https://doi.org/10.1093/acprof:oso/9780199213900.001.0001} {\emph {\bibinfo {title} {The Theory of Open Quantum Systems}}}\ (\bibinfo  {publisher} {Oxford University Press},\ \bibinfo {year} {2007})\BibitemShut {NoStop}%
\bibitem [{\citenamefont {Lindblad}(1976)}]{lindblad1976generators}%
  \BibitemOpen
  \bibfield  {author} {\bibinfo {author} {\bibfnamefont {G.}~\bibnamefont {Lindblad}},\ }\bibfield  {title} {\bibinfo {title} {On the generators of quantum dynamical semigroups},\ }\href {https://doi.org/10.1007/BF01608499} {\bibfield  {journal} {\bibinfo  {journal} {Communications in Mathematical Physics}\ }\textbf {\bibinfo {volume} {48}},\ \bibinfo {pages} {119} (\bibinfo {year} {1976})}\BibitemShut {NoStop}%
\bibitem [{\citenamefont {Gorini}\ \emph {et~al.}(1976)\citenamefont {Gorini}, \citenamefont {Kossakowski},\ and\ \citenamefont {Sudarshan}}]{gorini1976completely}%
  \BibitemOpen
  \bibfield  {author} {\bibinfo {author} {\bibfnamefont {V.}~\bibnamefont {Gorini}}, \bibinfo {author} {\bibfnamefont {A.}~\bibnamefont {Kossakowski}},\ and\ \bibinfo {author} {\bibfnamefont {E.~C.~G.}\ \bibnamefont {Sudarshan}},\ }\bibfield  {title} {\bibinfo {title} {Completely positive dynamical semigroups of {$N$}‐level systems},\ }\href {https://doi.org/10.1063/1.522979} {\bibfield  {journal} {\bibinfo  {journal} {Journal of Mathematical Physics}\ }\textbf {\bibinfo {volume} {17}},\ \bibinfo {pages} {821} (\bibinfo {year} {1976})}\BibitemShut {NoStop}%
\bibitem [{\citenamefont {Yoshihara}\ \emph {et~al.}(2017{\natexlab{a}})\citenamefont {Yoshihara}, \citenamefont {Fuse}, \citenamefont {Ashhab}, \citenamefont {Kakuyanagi}, \citenamefont {Saito},\ and\ \citenamefont {Semba}}]{yoshihara2017natphys}%
  \BibitemOpen
  \bibfield  {author} {\bibinfo {author} {\bibfnamefont {F.}~\bibnamefont {Yoshihara}}, \bibinfo {author} {\bibfnamefont {T.}~\bibnamefont {Fuse}}, \bibinfo {author} {\bibfnamefont {S.}~\bibnamefont {Ashhab}}, \bibinfo {author} {\bibfnamefont {K.}~\bibnamefont {Kakuyanagi}}, \bibinfo {author} {\bibfnamefont {S.}~\bibnamefont {Saito}},\ and\ \bibinfo {author} {\bibfnamefont {K.}~\bibnamefont {Semba}},\ }\bibfield  {title} {\bibinfo {title} {Superconducting qubit--oscillator circuit beyond the ultrastrong-coupling regime},\ }\href {https://doi.org/10.1038/nphys3906} {\bibfield  {journal} {\bibinfo  {journal} {Nature Physics}\ }\textbf {\bibinfo {volume} {13}},\ \bibinfo {pages} {44} (\bibinfo {year} {2017}{\natexlab{a}})}\BibitemShut {NoStop}%
\bibitem [{\citenamefont {Yoshihara}\ \emph {et~al.}(2017{\natexlab{b}})\citenamefont {Yoshihara}, \citenamefont {Fuse}, \citenamefont {Ashhab}, \citenamefont {Kakuyanagi}, \citenamefont {Saito},\ and\ \citenamefont {Semba}}]{yoshihara2017pra}%
  \BibitemOpen
  \bibfield  {author} {\bibinfo {author} {\bibfnamefont {F.}~\bibnamefont {Yoshihara}}, \bibinfo {author} {\bibfnamefont {T.}~\bibnamefont {Fuse}}, \bibinfo {author} {\bibfnamefont {S.}~\bibnamefont {Ashhab}}, \bibinfo {author} {\bibfnamefont {K.}~\bibnamefont {Kakuyanagi}}, \bibinfo {author} {\bibfnamefont {S.}~\bibnamefont {Saito}},\ and\ \bibinfo {author} {\bibfnamefont {K.}~\bibnamefont {Semba}},\ }\bibfield  {title} {\bibinfo {title} {Characteristic spectra of circuit quantum electrodynamics systems from the ultrastrong- to the deep-strong-coupling regime},\ }\href {https://doi.org/10.1103/PhysRevA.95.053824} {\bibfield  {journal} {\bibinfo  {journal} {Phys. Rev. A}\ }\textbf {\bibinfo {volume} {95}},\ \bibinfo {pages} {053824} (\bibinfo {year} {2017}{\natexlab{b}})}\BibitemShut {NoStop}%
\bibitem [{\citenamefont {Deng}\ \emph {et~al.}(2015)\citenamefont {Deng}, \citenamefont {Orgiazzi}, \citenamefont {Shen}, \citenamefont {Ashhab},\ and\ \citenamefont {Lupascu}}]{deng2015prl}%
  \BibitemOpen
  \bibfield  {author} {\bibinfo {author} {\bibfnamefont {C.}~\bibnamefont {Deng}}, \bibinfo {author} {\bibfnamefont {J.-L.}\ \bibnamefont {Orgiazzi}}, \bibinfo {author} {\bibfnamefont {F.}~\bibnamefont {Shen}}, \bibinfo {author} {\bibfnamefont {S.}~\bibnamefont {Ashhab}},\ and\ \bibinfo {author} {\bibfnamefont {A.}~\bibnamefont {Lupascu}},\ }\bibfield  {title} {\bibinfo {title} {Observation of floquet states in a strongly driven artificial atom},\ }\href {https://doi.org/10.1103/PhysRevLett.115.133601} {\bibfield  {journal} {\bibinfo  {journal} {Phys. Rev. Lett.}\ }\textbf {\bibinfo {volume} {115}},\ \bibinfo {pages} {133601} (\bibinfo {year} {2015})}\BibitemShut {NoStop}%
\bibitem [{\citenamefont {Niemczyk}\ \emph {et~al.}(2010)\citenamefont {Niemczyk}, \citenamefont {Deppe}, \citenamefont {Huebl}, \citenamefont {Menzel}, \citenamefont {Hocke}, \citenamefont {Schwarz}, \citenamefont {Garcia-Ripoll}, \citenamefont {Zueco}, \citenamefont {H{\"u}mmer}, \citenamefont {Solano}, \citenamefont {Marx},\ and\ \citenamefont {Gross}}]{Niemczyk2010}%
  \BibitemOpen
  \bibfield  {author} {\bibinfo {author} {\bibfnamefont {T.}~\bibnamefont {Niemczyk}}, \bibinfo {author} {\bibfnamefont {F.}~\bibnamefont {Deppe}}, \bibinfo {author} {\bibfnamefont {H.}~\bibnamefont {Huebl}}, \bibinfo {author} {\bibfnamefont {E.~P.}\ \bibnamefont {Menzel}}, \bibinfo {author} {\bibfnamefont {F.}~\bibnamefont {Hocke}}, \bibinfo {author} {\bibfnamefont {M.~J.}\ \bibnamefont {Schwarz}}, \bibinfo {author} {\bibfnamefont {J.~J.}\ \bibnamefont {Garcia-Ripoll}}, \bibinfo {author} {\bibfnamefont {D.}~\bibnamefont {Zueco}}, \bibinfo {author} {\bibfnamefont {T.}~\bibnamefont {H{\"u}mmer}}, \bibinfo {author} {\bibfnamefont {E.}~\bibnamefont {Solano}}, \bibinfo {author} {\bibfnamefont {A.}~\bibnamefont {Marx}},\ and\ \bibinfo {author} {\bibfnamefont {R.}~\bibnamefont {Gross}},\ }\bibfield  {title} {\bibinfo {title} {Circuit quantum electrodynamics in the ultrastrong-coupling regime},\ }\href {https://doi.org/10.1038/nphys1730} {\bibfield  {journal} {\bibinfo  {journal} {Nature Physics}\ }\textbf {\bibinfo
  {volume} {6}},\ \bibinfo {pages} {772} (\bibinfo {year} {2010})}\BibitemShut {NoStop}%
\bibitem [{\citenamefont {Fink}\ \emph {et~al.}(2008)\citenamefont {Fink}, \citenamefont {G{\"o}ppl}, \citenamefont {Baur}, \citenamefont {Bianchetti}, \citenamefont {Leek}, \citenamefont {Blais},\ and\ \citenamefont {Wallraff}}]{Fink2008}%
  \BibitemOpen
  \bibfield  {author} {\bibinfo {author} {\bibfnamefont {J.~M.}\ \bibnamefont {Fink}}, \bibinfo {author} {\bibfnamefont {M.}~\bibnamefont {G{\"o}ppl}}, \bibinfo {author} {\bibfnamefont {M.}~\bibnamefont {Baur}}, \bibinfo {author} {\bibfnamefont {R.}~\bibnamefont {Bianchetti}}, \bibinfo {author} {\bibfnamefont {P.~J.}\ \bibnamefont {Leek}}, \bibinfo {author} {\bibfnamefont {A.}~\bibnamefont {Blais}},\ and\ \bibinfo {author} {\bibfnamefont {A.}~\bibnamefont {Wallraff}},\ }\bibfield  {title} {\bibinfo {title} {Climbing the {J}aynes--{C}ummings ladder and observing its nonlinearity in a cavity {QED} system},\ }\href {https://doi.org/10.1038/nature07112} {\bibfield  {journal} {\bibinfo  {journal} {Nature}\ }\textbf {\bibinfo {volume} {454}},\ \bibinfo {pages} {315} (\bibinfo {year} {2008})}\BibitemShut {NoStop}%
\bibitem [{\citenamefont {Yoshikawa}\ \emph {et~al.}(2013)\citenamefont {Yoshikawa}, \citenamefont {Makino}, \citenamefont {Kurata}, \citenamefont {van Loock},\ and\ \citenamefont {Furusawa}}]{Yoshikawa2013prx}%
  \BibitemOpen
  \bibfield  {author} {\bibinfo {author} {\bibfnamefont {J.-i.}\ \bibnamefont {Yoshikawa}}, \bibinfo {author} {\bibfnamefont {K.}~\bibnamefont {Makino}}, \bibinfo {author} {\bibfnamefont {S.}~\bibnamefont {Kurata}}, \bibinfo {author} {\bibfnamefont {P.}~\bibnamefont {van Loock}},\ and\ \bibinfo {author} {\bibfnamefont {A.}~\bibnamefont {Furusawa}},\ }\bibfield  {title} {\bibinfo {title} {Creation, storage, and on-demand release of optical quantum states with a negative wigner function},\ }\href {https://doi.org/10.1103/PhysRevX.3.041028} {\bibfield  {journal} {\bibinfo  {journal} {Phys. Rev. X}\ }\textbf {\bibinfo {volume} {3}},\ \bibinfo {pages} {041028} (\bibinfo {year} {2013})}\BibitemShut {NoStop}%
\bibitem [{\citenamefont {Lau}\ and\ \citenamefont {Plenio}(2016)}]{lau2016PhysRevLett}%
  \BibitemOpen
  \bibfield  {author} {\bibinfo {author} {\bibfnamefont {H.-K.}\ \bibnamefont {Lau}}\ and\ \bibinfo {author} {\bibfnamefont {M.~B.}\ \bibnamefont {Plenio}},\ }\bibfield  {title} {\bibinfo {title} {Universal quantum computing with arbitrary continuous-variable encoding},\ }\href {https://doi.org/10.1103/PhysRevLett.117.100501} {\bibfield  {journal} {\bibinfo  {journal} {Phys. Rev. Lett.}\ }\textbf {\bibinfo {volume} {117}},\ \bibinfo {pages} {100501} (\bibinfo {year} {2016})}\BibitemShut {NoStop}%
\bibitem [{\citenamefont {Lee}\ \emph {et~al.}(2024)\citenamefont {Lee}, \citenamefont {Kang}, \citenamefont {Lee}, \citenamefont {Jeong}, \citenamefont {Jiang},\ and\ \citenamefont {Lee}}]{lee2024prxquantum}%
  \BibitemOpen
  \bibfield  {author} {\bibinfo {author} {\bibfnamefont {J.}~\bibnamefont {Lee}}, \bibinfo {author} {\bibfnamefont {N.}~\bibnamefont {Kang}}, \bibinfo {author} {\bibfnamefont {S.-H.}\ \bibnamefont {Lee}}, \bibinfo {author} {\bibfnamefont {H.}~\bibnamefont {Jeong}}, \bibinfo {author} {\bibfnamefont {L.}~\bibnamefont {Jiang}},\ and\ \bibinfo {author} {\bibfnamefont {S.-W.}\ \bibnamefont {Lee}},\ }\bibfield  {title} {\bibinfo {title} {Fault-tolerant quantum computation by hybrid qubits with bosonic cat code and single photons},\ }\href {https://doi.org/10.1103/PRXQuantum.5.030322} {\bibfield  {journal} {\bibinfo  {journal} {PRX Quantum}\ }\textbf {\bibinfo {volume} {5}},\ \bibinfo {pages} {030322} (\bibinfo {year} {2024})}\BibitemShut {NoStop}%
\bibitem [{\citenamefont {Xu}\ \emph {et~al.}(2023)\citenamefont {Xu}, \citenamefont {Zheng}, \citenamefont {Wang}, \citenamefont {Zoller}, \citenamefont {Clerk},\ and\ \citenamefont {Jiang}}]{xu2023npjquantuminfo}%
  \BibitemOpen
  \bibfield  {author} {\bibinfo {author} {\bibfnamefont {Q.}~\bibnamefont {Xu}}, \bibinfo {author} {\bibfnamefont {G.}~\bibnamefont {Zheng}}, \bibinfo {author} {\bibfnamefont {Y.-X.}\ \bibnamefont {Wang}}, \bibinfo {author} {\bibfnamefont {P.}~\bibnamefont {Zoller}}, \bibinfo {author} {\bibfnamefont {A.~A.}\ \bibnamefont {Clerk}},\ and\ \bibinfo {author} {\bibfnamefont {L.}~\bibnamefont {Jiang}},\ }\bibfield  {title} {\bibinfo {title} {Autonomous quantum error correction and fault-tolerant quantum computation with squeezed cat qubits},\ }\href {https://doi.org/10.1038/s41534-023-00746-0} {\bibfield  {journal} {\bibinfo  {journal} {npj Quantum Information}\ }\textbf {\bibinfo {volume} {9}},\ \bibinfo {pages} {78} (\bibinfo {year} {2023})}\BibitemShut {NoStop}%
\bibitem [{\citenamefont {Paparelle}\ \emph {et~al.}(2025)\citenamefont {Paparelle}, \citenamefont {Mousavi}, \citenamefont {Scazza}, \citenamefont {Bassi}, \citenamefont {Paris},\ and\ \citenamefont {Zavatta}}]{paparelle2025optexpress}%
  \BibitemOpen
  \bibfield  {author} {\bibinfo {author} {\bibfnamefont {I.}~\bibnamefont {Paparelle}}, \bibinfo {author} {\bibfnamefont {F.}~\bibnamefont {Mousavi}}, \bibinfo {author} {\bibfnamefont {F.}~\bibnamefont {Scazza}}, \bibinfo {author} {\bibfnamefont {A.}~\bibnamefont {Bassi}}, \bibinfo {author} {\bibfnamefont {M.}~\bibnamefont {Paris}},\ and\ \bibinfo {author} {\bibfnamefont {A.}~\bibnamefont {Zavatta}},\ }\bibfield  {title} {\bibinfo {title} {Experimental direct quantum communication with squeezed states},\ }\href {https://doi.org/10.1364/OE.538593} {\bibfield  {journal} {\bibinfo  {journal} {Opt. Express}\ }\textbf {\bibinfo {volume} {33}},\ \bibinfo {pages} {28917} (\bibinfo {year} {2025})}\BibitemShut {NoStop}%
\bibitem [{\citenamefont {Reiserer}\ and\ \citenamefont {Rempe}(2015)}]{Reiserer2015RevModPhys}%
  \BibitemOpen
  \bibfield  {author} {\bibinfo {author} {\bibfnamefont {A.}~\bibnamefont {Reiserer}}\ and\ \bibinfo {author} {\bibfnamefont {G.}~\bibnamefont {Rempe}},\ }\bibfield  {title} {\bibinfo {title} {Cavity-based quantum networks with single atoms and optical photons},\ }\href {https://doi.org/10.1103/RevModPhys.87.1379} {\bibfield  {journal} {\bibinfo  {journal} {Rev. Mod. Phys.}\ }\textbf {\bibinfo {volume} {87}},\ \bibinfo {pages} {1379} (\bibinfo {year} {2015})}\BibitemShut {NoStop}%
\bibitem [{\citenamefont {Schnabel}(2017)}]{schnabel2017physrep}%
  \BibitemOpen
  \bibfield  {author} {\bibinfo {author} {\bibfnamefont {R.}~\bibnamefont {Schnabel}},\ }\bibfield  {title} {\bibinfo {title} {Squeezed states of light and their applications in laser interferometers},\ }\href {https://doi.org/https://doi.org/10.1016/j.physrep.2017.04.001} {\bibfield  {journal} {\bibinfo  {journal} {Physics Reports}\ }\textbf {\bibinfo {volume} {684}},\ \bibinfo {pages} {1} (\bibinfo {year} {2017})}\BibitemShut {NoStop}%
\bibitem [{\citenamefont {Mart{\'i}nez-Pe{\~{n}}a}\ \emph {et~al.}(2023)\citenamefont {Mart{\'i}nez-Pe{\~{n}}a}, \citenamefont {Nokkala}, \citenamefont {Giorgi}, \citenamefont {Zambrini},\ and\ \citenamefont {Soriano}}]{MartinezPena2023}%
  \BibitemOpen
  \bibfield  {author} {\bibinfo {author} {\bibfnamefont {R.}~\bibnamefont {Mart{\'i}nez-Pe{\~{n}}a}}, \bibinfo {author} {\bibfnamefont {J.}~\bibnamefont {Nokkala}}, \bibinfo {author} {\bibfnamefont {G.~L.}\ \bibnamefont {Giorgi}}, \bibinfo {author} {\bibfnamefont {R.}~\bibnamefont {Zambrini}},\ and\ \bibinfo {author} {\bibfnamefont {M.~C.}\ \bibnamefont {Soriano}},\ }\bibfield  {title} {\bibinfo {title} {Information processing capacity of spin-based quantum reservoir computing systems},\ }\href {https://doi.org/10.1007/s12559-020-09772-y} {\bibfield  {journal} {\bibinfo  {journal} {Cognitive Computation}\ }\textbf {\bibinfo {volume} {15}},\ \bibinfo {pages} {1440} (\bibinfo {year} {2023})}\BibitemShut {NoStop}%
\bibitem [{\citenamefont {Sannia}\ \emph {et~al.}(2024)\citenamefont {Sannia}, \citenamefont {Mart{\'{i}}nez-Pe{\~{n}}a}, \citenamefont {Soriano}, \citenamefont {Giorgi},\ and\ \citenamefont {Zambrini}}]{Sannia2024}%
  \BibitemOpen
  \bibfield  {author} {\bibinfo {author} {\bibfnamefont {A.}~\bibnamefont {Sannia}}, \bibinfo {author} {\bibfnamefont {R.}~\bibnamefont {Mart{\'{i}}nez-Pe{\~{n}}a}}, \bibinfo {author} {\bibfnamefont {M.~C.}\ \bibnamefont {Soriano}}, \bibinfo {author} {\bibfnamefont {G.~L.}\ \bibnamefont {Giorgi}},\ and\ \bibinfo {author} {\bibfnamefont {R.}~\bibnamefont {Zambrini}},\ }\bibfield  {title} {\bibinfo {title} {Dissipation as a resource for {Q}uantum {R}eservoir {C}omputing},\ }\href {https://doi.org/10.22331/q-2024-03-20-1291} {\bibfield  {journal} {\bibinfo  {journal} {{Quantum}}\ }\textbf {\bibinfo {volume} {8}},\ \bibinfo {pages} {1291} (\bibinfo {year} {2024})}\BibitemShut {NoStop}%
\bibitem [{\citenamefont {Zhu}\ \emph {et~al.}(2025)\citenamefont {Zhu}, \citenamefont {Ehlers}, \citenamefont {Nurdin},\ and\ \citenamefont {Soh}}]{Zhu2025}%
  \BibitemOpen
  \bibfield  {author} {\bibinfo {author} {\bibfnamefont {C.}~\bibnamefont {Zhu}}, \bibinfo {author} {\bibfnamefont {P.~J.}\ \bibnamefont {Ehlers}}, \bibinfo {author} {\bibfnamefont {H.~I.}\ \bibnamefont {Nurdin}},\ and\ \bibinfo {author} {\bibfnamefont {D.}~\bibnamefont {Soh}},\ }\bibfield  {title} {\bibinfo {title} {Minimalistic and scalable quantum reservoir computing enhanced with feedback},\ }\href {https://doi.org/10.1038/s41534-025-01144-4} {\bibfield  {journal} {\bibinfo  {journal} {npj Quantum Information}\ }\textbf {\bibinfo {volume} {11}},\ \bibinfo {pages} {195} (\bibinfo {year} {2025})}\BibitemShut {NoStop}%
\bibitem [{\citenamefont {Das}\ \emph {et~al.}(2026)\citenamefont {Das}, \citenamefont {Giorgi},\ and\ \citenamefont {Zambrini}}]{Das2026}%
  \BibitemOpen
  \bibfield  {author} {\bibinfo {author} {\bibfnamefont {S.}~\bibnamefont {Das}}, \bibinfo {author} {\bibfnamefont {G.~L.}\ \bibnamefont {Giorgi}},\ and\ \bibinfo {author} {\bibfnamefont {R.}~\bibnamefont {Zambrini}},\ }\bibfield  {title} {\bibinfo {title} {Quantum reservoir computing in {J}aynes-{C}ummings models: Nonlinear memory and time-series prediction},\ }\href {https://doi.org/10.1103/ffd3-ytbt} {\bibfield  {journal} {\bibinfo  {journal} {Phys. Rev. Res.}\ }\textbf {\bibinfo {volume} {8}},\ \bibinfo {pages} {023148} (\bibinfo {year} {2026})}\BibitemShut {NoStop}%
\bibitem [{\citenamefont {Tumbiolo}\ \emph {et~al.}(2026)\citenamefont {Tumbiolo}, \citenamefont {Maccone}, \citenamefont {Macchiavello}, \citenamefont {Paris},\ and\ \citenamefont {Guarnieri}}]{tumbiolo2026prl}%
  \BibitemOpen
  \bibfield  {author} {\bibinfo {author} {\bibfnamefont {E.}~\bibnamefont {Tumbiolo}}, \bibinfo {author} {\bibfnamefont {L.}~\bibnamefont {Maccone}}, \bibinfo {author} {\bibfnamefont {C.}~\bibnamefont {Macchiavello}}, \bibinfo {author} {\bibfnamefont {M.~G.~A.}\ \bibnamefont {Paris}},\ and\ \bibinfo {author} {\bibfnamefont {G.}~\bibnamefont {Guarnieri}},\ }\bibfield  {title} {\bibinfo {title} {Shake before use: Universal enhancement of quantum thermometry by unitary driving},\ }\href {https://doi.org/10.1103/z6rz-d9sq} {\bibfield  {journal} {\bibinfo  {journal} {Phys. Rev. Lett.}\ } (\bibinfo {year} {2026})}\BibitemShut {NoStop}%
\bibitem [{\citenamefont {Vahlbruch}\ \emph {et~al.}(2016)\citenamefont {Vahlbruch}, \citenamefont {Mehmet}, \citenamefont {Danzmann},\ and\ \citenamefont {Schnabel}}]{Vahlbruch2015prl}%
  \BibitemOpen
  \bibfield  {author} {\bibinfo {author} {\bibfnamefont {H.}~\bibnamefont {Vahlbruch}}, \bibinfo {author} {\bibfnamefont {M.}~\bibnamefont {Mehmet}}, \bibinfo {author} {\bibfnamefont {K.}~\bibnamefont {Danzmann}},\ and\ \bibinfo {author} {\bibfnamefont {R.}~\bibnamefont {Schnabel}},\ }\bibfield  {title} {\bibinfo {title} {Detection of 15 db squeezed states of light and their application for the absolute calibration of photoelectric quantum efficiency},\ }\href {https://doi.org/10.1103/PhysRevLett.117.110801} {\bibfield  {journal} {\bibinfo  {journal} {Phys. Rev. Lett.}\ }\textbf {\bibinfo {volume} {117}},\ \bibinfo {pages} {110801} (\bibinfo {year} {2016})}\BibitemShut {NoStop}%
\bibitem [{\citenamefont {Dassonneville}\ \emph {et~al.}(2021)\citenamefont {Dassonneville}, \citenamefont {Assouly}, \citenamefont {Peronnin}, \citenamefont {Clerk}, \citenamefont {Bienfait},\ and\ \citenamefont {Huard}}]{dassonneville2021PRXQuantum}%
  \BibitemOpen
  \bibfield  {author} {\bibinfo {author} {\bibfnamefont {R.}~\bibnamefont {Dassonneville}}, \bibinfo {author} {\bibfnamefont {R.}~\bibnamefont {Assouly}}, \bibinfo {author} {\bibfnamefont {T.}~\bibnamefont {Peronnin}}, \bibinfo {author} {\bibfnamefont {A.}~\bibnamefont {Clerk}}, \bibinfo {author} {\bibfnamefont {A.}~\bibnamefont {Bienfait}},\ and\ \bibinfo {author} {\bibfnamefont {B.}~\bibnamefont {Huard}},\ }\bibfield  {title} {\bibinfo {title} {Dissipative stabilization of squeezing beyond 3 db in a microwave mode},\ }\href {https://doi.org/10.1103/PRXQuantum.2.020323} {\bibfield  {journal} {\bibinfo  {journal} {PRX Quantum}\ }\textbf {\bibinfo {volume} {2}},\ \bibinfo {pages} {020323} (\bibinfo {year} {2021})}\BibitemShut {NoStop}%
\bibitem [{\citenamefont {Eickbusch}\ \emph {et~al.}(2022)\citenamefont {Eickbusch}, \citenamefont {Sivak}, \citenamefont {Ding}, \citenamefont {Elder}, \citenamefont {Jha}, \citenamefont {Venkatraman}, \citenamefont {Royer}, \citenamefont {Girvin}, \citenamefont {Schoelkopf},\ and\ \citenamefont {Devoret}}]{eickbusch2022natphys}%
  \BibitemOpen
  \bibfield  {author} {\bibinfo {author} {\bibfnamefont {A.}~\bibnamefont {Eickbusch}}, \bibinfo {author} {\bibfnamefont {V.}~\bibnamefont {Sivak}}, \bibinfo {author} {\bibfnamefont {A.~Z.}\ \bibnamefont {Ding}}, \bibinfo {author} {\bibfnamefont {S.~S.}\ \bibnamefont {Elder}}, \bibinfo {author} {\bibfnamefont {S.~R.}\ \bibnamefont {Jha}}, \bibinfo {author} {\bibfnamefont {J.}~\bibnamefont {Venkatraman}}, \bibinfo {author} {\bibfnamefont {B.}~\bibnamefont {Royer}}, \bibinfo {author} {\bibfnamefont {S.~M.}\ \bibnamefont {Girvin}}, \bibinfo {author} {\bibfnamefont {R.~J.}\ \bibnamefont {Schoelkopf}},\ and\ \bibinfo {author} {\bibfnamefont {M.~H.}\ \bibnamefont {Devoret}},\ }\bibfield  {title} {\bibinfo {title} {Fast universal control of an oscillator with weak dispersive coupling to a qubit},\ }\href {https://doi.org/10.1038/s41567-022-01776-9} {\bibfield  {journal} {\bibinfo  {journal} {Nature Physics}\ }\textbf {\bibinfo {volume} {18}},\ \bibinfo {pages} {1464} (\bibinfo {year} {2022})}\BibitemShut {NoStop}%
\bibitem [{\citenamefont {Cai}\ \emph {et~al.}(2025)\citenamefont {Cai}, \citenamefont {Deng}, \citenamefont {Zhang}, \citenamefont {Ni}, \citenamefont {Mai}, \citenamefont {Huang}, \citenamefont {Zheng}, \citenamefont {Hu}, \citenamefont {Liu}, \citenamefont {Xu},\ and\ \citenamefont {Yu}}]{cai2025natcomm}%
  \BibitemOpen
  \bibfield  {author} {\bibinfo {author} {\bibfnamefont {Y.}~\bibnamefont {Cai}}, \bibinfo {author} {\bibfnamefont {X.}~\bibnamefont {Deng}}, \bibinfo {author} {\bibfnamefont {L.}~\bibnamefont {Zhang}}, \bibinfo {author} {\bibfnamefont {Z.}~\bibnamefont {Ni}}, \bibinfo {author} {\bibfnamefont {J.}~\bibnamefont {Mai}}, \bibinfo {author} {\bibfnamefont {P.}~\bibnamefont {Huang}}, \bibinfo {author} {\bibfnamefont {P.}~\bibnamefont {Zheng}}, \bibinfo {author} {\bibfnamefont {L.}~\bibnamefont {Hu}}, \bibinfo {author} {\bibfnamefont {S.}~\bibnamefont {Liu}}, \bibinfo {author} {\bibfnamefont {Y.}~\bibnamefont {Xu}},\ and\ \bibinfo {author} {\bibfnamefont {D.}~\bibnamefont {Yu}},\ }\bibfield  {title} {\bibinfo {title} {Quantum squeezing amplification with a weak kerr nonlinear oscillator},\ }\href {https://doi.org/10.1038/s41467-025-67699-0} {\bibfield  {journal} {\bibinfo  {journal} {Nature Communications}\ }\textbf {\bibinfo {volume} {17}},\ \bibinfo {pages} {970} (\bibinfo {year} {2025})}\BibitemShut {NoStop}%
\end{thebibliography}%

\onecolumngrid

\newpage 
\section*{End Matter}    
\appendix
\twocolumngrid
\setcounter{equation}{0}
\renewcommand{\thesection}{A}

\textit{Appendix A: Squeezed state generation.}---To illustrate how our  optimal control framework generalizes to other nonclassical states, we focus on the preparation of single-mode squeezed vacuum states of the cavity field, defined as $S(r,\theta)\ket{0}$, where
\begin{equation}
    S(r,\theta)=\exp\left[\frac{1}{2}r\left(e^{-i\theta} a^2 - e^{i\theta} a^{\dagger 2}\right)\right]
\end{equation}
is the squeezing operator, 
$r \geq 0$ the squeezing strength, and $\theta$ the squeezing angle. Squeezed states are characterized by reduced fluctuations in one field quadrature at the expense of increased fluctuations in the conjugate one. For a single-mode squeezed vacuum, the variance of the squeezed quadrature is
$  \Delta X^2 = e^{-2r}/2$, where $\Delta X_0^2 = 1/2$ is the vacuum variance.
In experiments, squeezing is typically quantified in decibels (dB) as
\begin{equation}
    r_{\mathrm{dB}} = -10 \log_{10}\left( \frac{\Delta X^2}{\Delta X_0^2} \right)
    = 20\,r\,\log_{10}(e),
\end{equation}
which can be inverted as $r \simeq r_{\mathrm{dB}}/8.686$.

\begin{figure*}[!ht]
    \centering
    \includegraphics[width=\linewidth]{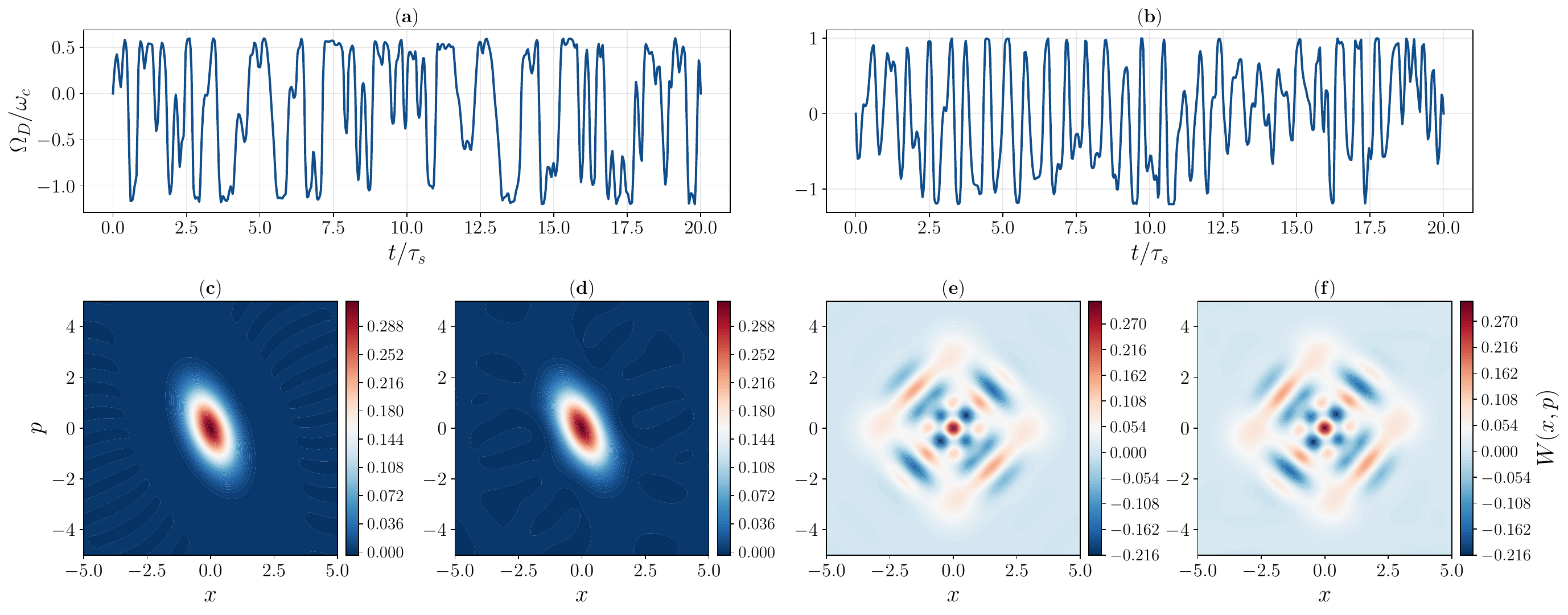}
    \caption{
    Optimal control strategy for the generation of squeezed and cat states.
    (a, b) Optimized control drives $\Omega_D(t)/\omega_c$ used for the generation of (a) the squeezed vacuum state and (b) the superposition of Schr\"odinger-cat states in Eq.~\eqref{eq:traget_superpos_cat_states}. 
    (c--f) Wigner functions of the cavity field comparing the ideal target states (c, e) with the final states obtained via optimal control (d, f). Specifically, (c, d) refer to the squeezed vacuum state with squeezing strength $r=3\,$dB and squeezing angle $\theta=\pi/4$, while (e, f) refer to the Schr\"odinger-cat state with $\alpha=2$.
    The number of time intervals for the GRAPE refinement is $N_t=400$, 
    the other parameters are the same as those used in Fig.~\ref{fig:drive}.
    }
    \label{fig:cat_squeezed_tot}
\end{figure*}

To determine the optimal control protocol, we employ the same control parametrization, $\Omega_D(t)$, and multi-stage optimization pipeline (CRAB, GRAPE, and final interpolation) developed for Fock-state generation, where the infidelity cost functional is now defined with respect to the target squeezed state. As a first benchmark, we focus on the generation of a mildly squeezed state ($r_{\rm dB}=3$) with a finite squeezing angle ($\theta = \pi/4$),  resulting in a rotated squeezing axis in phase space. The drive obtained from this optimization [Fig. \ref{fig:cat_squeezed_tot}(a)] prepares a state that achieves an exceptional fidelity of $F \simeq 0.999$ with respect to the target single-mode squeezed vacuum state, as reflected in the excellent agreement between the Wigner functions [Figs.~\ref{fig:cat_squeezed_tot}(c,d)].

\begin{figure*}[!ht]
    \centering
    \includegraphics[width=\linewidth]{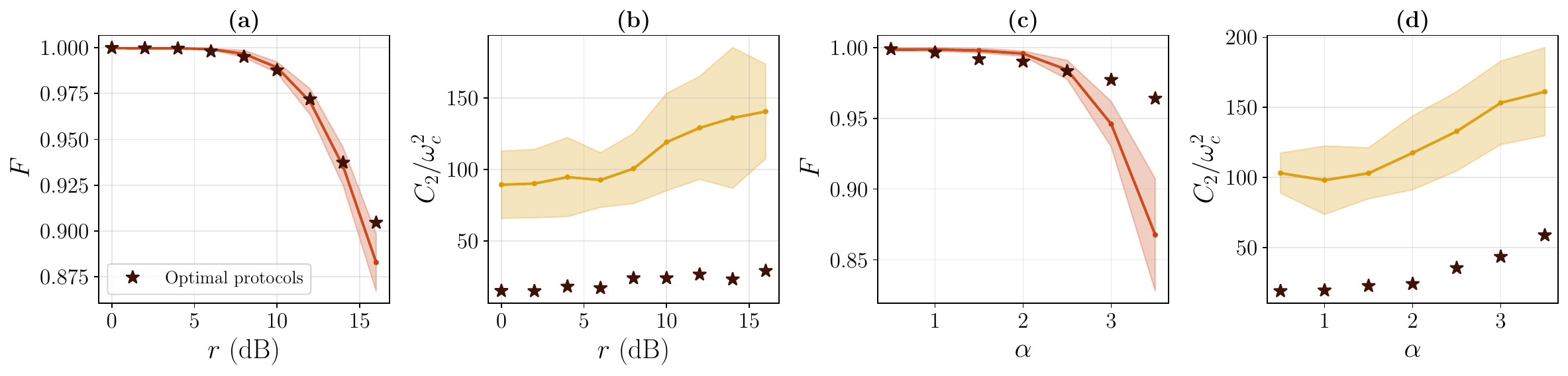}
    \caption{
    Performance scaling of squeezed and cat-state generation. 
    (a, b) Squeezed-vacuum state generation as a function of the squeezing strength $r_\mathrm{dB}$ at $\theta=0$.
    (c, d) Generation of Schr\"{o}dinger-cat-state superpositions as a function of $\alpha$ [cf. Eq.~\eqref{eq:traget_superpos_cat_states}].
    We show the average (a, c) fidelity, and (b, d) control cost $C_2$~\eqref{eq:cost} associated with the driving.
    Markers (mean value) and error bands (standard deviation) are obtained from 10 independent unconstrained CRAB optimizations with different random seeds.
    The star markers indicate the performance of the optimal protocols obtained by refining the best CRAB solutions through a subsequent GRAPE optimization followed by a final pulse-smoothing interpolation step.
    The number of time intervals for the GRAPE refinement is $N_t=600$ (for squeezed states) and $N_t=400$ (for cat states), 
    the other parameters are the same as those used in Fig.~\ref{fig:drive}.
     }
    \label{fig:scaling_tot}
\end{figure*}

Then, we investigate the protocol's performance across experimentally attainable squeezing strengths $r_{\mathrm{dB}} \in [0,16]$ \cite{Vahlbruch2015prl,dassonneville2021PRXQuantum,eickbusch2022natphys,cai2025natcomm}, setting $\theta=0$ for simplicity.
For each squeezing strength $r$, we run 10 independent bare CRAB optimizations (imposing no constraints on the control amplitudes) with different random seeds to evaluate the average fidelity and control cost [Fig.~\ref{fig:scaling_tot}(a,b)].
For weak to moderate squeezing, the protocol consistently yields high fidelities ($F \gtrsim 0.98$), performing comparably to Fock-state generation. At stronger squeezing, the average fidelity gradually degrades as $r$ increases, accompanied by larger fluctuations across different optimization trials that suggest an increasing complexity of the control landscape. Concurrently, the control cost is nearly constant in the weak squeezing regime, while increasing mildly at stronger squeezing. However, because this trend remains comparable to the statistical fluctuations across different realizations, no definitive scaling can be resolved within the error bars.

Fig.~\ref{fig:scaling_tot}(a,b) also shows the results obtained by selecting the best CRAB solutions (in terms of fidelity) for each $r$ and refining them via GRAPE under explicit amplitude bounds, followed by pulse smoothing. The refined fidelity slightly outperforms the bare CRAB average for all $r$, while the corresponding control cost is drastically reduced.

Overall, while no definitive scaling of the required control resources emerges, the results suggest that strongly squeezed states demand more structured control fields and exhibit larger variability in the optimization outcomes. Nevertheless, the protocol remains effective across the explored range of squeezing, demonstrating its versatility for nonclassical state generation. 

\renewcommand{\thesection}{B}

\textit{Appendix B: Generation of Schr\"odinger-cat-state superpositions}.---We further assess the versatility of our protocol by targeting superpositions of Schr\"odinger-cat states. 
The standard two-component even ($+$) and odd ($-$) Schr\"odinger-cat states are defined as
\begin{equation}
    \ket{C_{\alpha}^{\pm}} \propto  \ket{\alpha} \pm \ket{-\alpha},
\end{equation}
where $\ket{\alpha}$ denotes a coherent state, satisfying $a\ket{\alpha} = \alpha\ket{\alpha}$, with $\alpha \in \mathbb{C}$. 
The complex parameter $\alpha$ represents the displacement in the phase space, and its squared magnitude $|\alpha|^2$ corresponds to the average number of photons in each coherent component.

As an illustrative and more challenging benchmark, we test the protocol for the generation of a superposition of cat states, constructed along two orthogonal axes of the phase space: 
\begin{equation}
    \ket{\psi_\mathrm{target}} = \mathcal{N} \left( \ket{C_{\alpha}^+} - \ket{C_{i\alpha}^+} \right),
    \label{eq:traget_superpos_cat_states}
\end{equation}
where $\mathcal{N}$ is the appropriate normalization constant, and restricting to $\alpha \in \mathbb{R}$. This specific state features a highly non-classical checkered pattern of positive and negative interference fringes at the center of the Wigner function, making it extremely sensitive to phase-space displacements.
We employ the same control parametrization, $\Omega_D(t)$, and multi-stage optimization pipeline (CRAB, GRAPE, and final interpolation) developed for Fock and squeezed-state generation, where the infidelity cost functional is now defined with respect to the target cat state. 

Specifically, targeting the superposition state with $\alpha = 2$ (where each coherent component carries an average of $|\alpha|^2 = 4$ photons), the protocol achieves a remarkably high final fidelity of $F = 0.978$; the corresponding optimized drive and Wigner functions are presented in Fig.~\ref{fig:cat_squeezed_tot}(b,e,f), showcasing an almost perfect agreement between the target [Fig.~\ref{fig:cat_squeezed_tot}(e)] and reconstructed [Fig.~\ref{fig:cat_squeezed_tot}(f)] states. We emphasize that the protocol achieves a significantly reduced time-to-solution compared to standard gate-based approaches (see SM for details). 

Finally, we investigate how the protocol's performance in generating the target state in Eq.~\eqref{eq:traget_superpos_cat_states} scales with $\alpha$. For each value of $\alpha$, we run 10 independent unconstrained CRAB optimizations with different random seeds, and evaluate the fidelity and control cost, reporting their mean value and standard deviation in Fig.~\ref{fig:scaling_tot}(c,d). The fidelity is consistently high ($F \gtrsim 0.95$) for $\alpha<3$, and degrades for larger $\alpha$ [Fig.~\ref{fig:scaling_tot}(c)]. The control cost shows an overall increasing trend as $\alpha$ increases [Fig.~\ref{fig:scaling_tot}(d)]. 

Fig.~\ref{fig:scaling_tot}(c,d) also shows the results obtained by selecting the best CRAB solutions (in terms of fidelity) for each $\alpha$ and refining them via GRAPE under explicit amplitude bounds, followed by pulse smoothing. Notably, while this refined fidelity remains very high at low $\alpha$, it significantly outperforms the bare CRAB average for $\alpha > 2$. Consistent with our observations for Fock and squeezed states, the corresponding control cost is drastically reduced.

\end{document}


\title{Supplemental Material for \\
``Engineering Nonclassical States via the Dynamical Casimir Effect''}

\author{Maristella Crotti}
\affiliation{\como}
\affiliation{\infnM}
\author{Luca Razzoli}
\affiliation{\pavia}
\affiliation{\infnP}
\author{Giacomo Guarnieri}
\affiliation{\pavia}
\affiliation{\infnP}
\author{Luigi Giannelli}
\affiliation{\catania}
\affiliation{\infnC}
\author{Giuseppe A. Falci}
\affiliation{\catania}
\affiliation{\infnC}
\author{Giuliano Benenti}
\affiliation{\como}
\affiliation{\infnM}

\begin{abstract}
    This supplemental material (SM) provides details on the optimal-control framework used to design the control fields for nonclassical cavity state generation. After detailing the specific implementations of the CRAB and GRAPE optimal control techniques, we present a numerical characterization of the control landscape for Fock-state generation. We then detail the dissipation model adopted for the open system dynamics, and conclude by elaborating on the experimental feasibility of our protocol. Throughout the SM, section, equation, and figure numbers prefixed with ``S'' refer to the SM, while those without such prefix (or with prefix ``A'' or ``B'') refer to the main text (End Matter).
\end{abstract}

\maketitle

\date{\today}

\section{Optimal control methods}
Our goal is to generate nonclassical \textit{cavity} states within the Rabi model [Eq. (1)] by leveraging an external drive which modulates the qubit's frequency [Eq. (2)].
To optimize the control field $\Omega_D(t)$ we employ two standard optimal control strategies widely used in the literature: the Chopped RAndom Basis (CRAB) method \cite{PhysRevA.84.022326,Muller2022} and the GRAPE (GRadient Ascent Pulse Engineering) method \cite{KHANEJA2005296}. 
While both methods share the objective of maximizing the fidelity with respect to a target state, they differ in the parametrization of the control field and in the numerical representation of the system dynamics; in our setting, this leads to slightly different implementations of the cost functional.

\subsection{CRAB optimization}
In the CRAB approach, the control field $\Omega_D(t)$ is parametrized in a truncated basis, thereby reducing the original infinite-dimensional control problem to a finite-dimensional optimization \cite{PhysRevA.84.022326,Muller2022}. The control is expressed as a finite Fourier-like expansion with randomized frequencies,
\begin{equation}
\Omega_D(t) = s(t)\sum_{k=1}^{N_f} \left[ A_k \sin(\omega_k t) + B_k \cos(\omega_k t) \right] ,
\label{eq:drive_expansion_CRAB}
\end{equation}
where $\{A_k, B_k\}$ are variational parameters to be optimized, and  $N_f$ is the number of harmonics.
Within the CRAB framework, the frequencies $\omega_k$ can be generated according to different prescriptions. In particular, they may be sampled uniformly from a
prescribed interval $[\omega_{\min}, \omega_{\max}]$, as done in the present work, or defined via a randomized Fourier construction,
\begin{equation}
\omega_k = \frac{2\pi}{T}(k + r_k), \quad r_k \in [-0.5, 0.5],
\end{equation}
where $T$ denotes the protocol duration, and $r_k$ are uniformly distributed random variables. This latter choice introduces stochasticity into the Fourier basis. 
The stochasticity introduced by either frequency-generation scheme
improves exploration of the control landscape, and helps avoid trapping in local minima associated with fixed spectral grids.
The envelope function $s(t)$ enforces smooth boundary conditions at the initial and final times, ensuring $\Omega_D(0)=\Omega_D(T)=0$. We adopt 
\begin{equation}
s(t) = \left(1 - e^{-(t/\sigma)^2}\right)\left(1 - e^{-((t-T)/\sigma)^2}\right),
\end{equation}
with switching parameter $\sigma=0.7/\omega_c$.

The system dynamics are obtained by numerically solving the time-dependent Schrödinger equation for the full cavity--qubit system, starting from the initial state $\rho(0) = \dyad{\psi_0}$ with $\ket{\psi_0} = \ket{0}\ket{g}$, using \texttt{QuTiP} \cite{Johansson2012,Johansson2013} and assuming closed-system evolution. At the final time, the reduced state of the cavity mode is obtained by tracing out the qubit degrees of freedom, $\rho_{\mathrm{cavity}}(T)=\mathrm{Tr}_{\mathrm{qubit}}[\rho(T)]$.

The optimization aims at minimizing a cost functional $\mathcal{C}[\Omega_D(t)]=1 - F$, defined as the infidelity with respect to the target cavity state.
The fidelity $F$ is computed between the reduced cavity state $\rho_{\mathrm{cavity}}(T)$
at the final time $T$ and the target Fock state $\ket{\psi_{\mathrm{target}}}$, i.e., 
\begin{equation}
F = \bra{\psi_{\mathrm{target}}} \rho_{\mathrm{cavity}}(T) \ket{\psi_{\mathrm{target}}}.
\label{eq:fidelity}
\end{equation}
The optimization is performed using the derivative-free Nelder--Mead simplex algorithm implemented in \texttt{SciPy} \cite{Gao2012, 2020SciPyNMeth}, which requires only evaluations of the cost functional without gradient information.

\subsection{GRAPE optimization} 
In contrast to CRAB-based approaches, GRAPE is a fully gradient-based optimal control method in which the control field is not expanded in a truncated basis but directly optimized on a fixed time grid \cite{KHANEJA2005296}.
The total evolution time $T$ is discretized into $N_t$ time steps of duration $\delta t = T/N_t$, and the control field is represented as a piecewise-constant function, $\Omega_D(t) \rightarrow \{\Omega_{D}^{(1)}, \Omega_{D}^{(2)}, \ldots, \Omega_{D}^{(N_t)}\}$, assumed constant within each interval.
At each time step $k$, the system evolves under the effective Hamiltonian $H_k = H_{R} + \Omega_{D}^{(k)} \sigma_z/2$ [see the Rabi Hamiltonian in Eq. (1) and the drive in Eq. (2)], with corresponding short-time propagator $U_k = \exp(-i H_k \delta t)$, and full evolution given by the time-ordered product $U(T) = U_{N_t} \cdots U_2 U_1$.
We propagate the full density matrix of the bipartite cavity--qubit system according to $\rho_{k+1} = U_k \rho_k U_k^\dagger$, starting from the initial state $\rho(0) = |\psi_0\rangle\langle \psi_0|$.

The control objective is defined in terms of the overlap with a target state embedded in the full Hilbert space of the cavity--qubit system,
\begin{equation}
F = \mathrm{Tr}\!\left[\rho_{\mathrm{target}} \, \rho(T)\right].
\label{eq:fidelity_grape}
\end{equation}
The objective is therefore the minimization of the cost function $\mathcal{C} = 1 - F$.
To compute gradients efficiently, we introduce an adjoint state propagated backward in time. Starting from $\lambda_{N_t} = \rho_{\mathrm{target}}$, the backward evolution is given by $\lambda_k = U_k^\dagger \lambda_{k+1} U_k$.
The gradient of the cost function with respect to each control amplitude is given by~\cite{KHANEJA2005296}
\begin{equation}
\frac{\partial \,\mathcal{C}}{\partial \Omega_{D}^{(k)}} 
=-\frac{\partial F}{\partial \Omega_{D}^{(k)}}
= \mathrm{Re}\left\{
\mathrm{Tr}\left[
\lambda_{k+1} \, i \delta t \,\left[\frac{\sigma_z}{2}, \rho_k\right]
\right]
\right\}.
\label{eq:grad}
\end{equation}
This expression provides an exact gradient of the cost function and avoids finite-difference evaluations, forming the basis of the GRAPE update rule.
In practice, the control amplitudes are optimized using a quasi-Newton optimization scheme based on the L-BFGS-B algorithm \cite{LBFGSB1995}. In this case, the optimization is formulated as a constrained finite-dimensional problem, where bounds on the control amplitudes are enforced as $\Omega_{\min} \leq \Omega_{D}^{(k)} \leq \Omega_{\max}$.
The L-BFGS-B implementation is supplied with both the cost function and its analytical gradient of Eq. \eqref{eq:grad}. This allows for efficient convergence while maintaining physical constraints on the control field.

For the reduced cavity state to be the pure target state $\ket{\psi_{\rm target}}$, the target cavity--qubit state employed in the GRAPE optimization must be separable, i.e.,
\begin{equation}
\rho_{\mathrm{target}} =
\begin{cases}
\dyad{\psi_{\mathrm{target}}} \otimes \dyad{g} & \text{if $\ket{\psi_\mathrm{target}}$ is even,}\\
\dyad{\psi_{\mathrm{target}}} \otimes \dyad{e} & \text{if $\ket{\psi_\mathrm{target}}$ is odd.}
\end{cases}
\label{eq:target}
\end{equation}
This conditional definition follows from the conservation of the total excitation parity operator $\Pi = \exp[i\pi (a^\dagger a + \sigma_+\sigma_-)]$ under the evolution generated by both the Rabi and the control Hamiltonians [Eqs. (1)--(2)], given the even-parity initial state $\ket{\psi_0}=\ket{0}\ket{g}$.\footnote{The eigenvalues of the total parity operator $\Pi$, $p=+1$ and $p=-1$, denote even and odd parity, respectively.}
In fact, parity conservation restricts the system dynamics to the even-parity subspace spanned by $\{\ket{2n}\ket{g},\ket{2n+1}\ket{e}\}_{n\in \mathbb{N}_0}$, and any target cavity--qubit state must belong to this subspace to be reachable, i.e., it must share the same total even parity as the initial state.
Therefore, the target cavity--qubit state in Eq. \eqref{eq:target} specializes as follows for the nonclassical states addressed in the present work.
For a target Fock state $\ket{\psi_{\rm target}} = \ket{n}$, an even (odd) photon number $n$ requires the qubit to be in the ground $\ket{g}$ (excited $\ket{e}$) state.
For a single-mode squeezed vacuum state $\ket{\psi_{\rm target}} = S(r,\theta)\ket{0}$, where $S(r,\theta)=\exp\left[\frac{1}{2}r\left(e^{-i\theta} a^2 - e^{i\theta} a^{\dagger 2}\right)\right]$, the qubit must be in the ground state $\ket{g}$. This is because $S(r,\theta)\ket{0}$ shares the same (bosonic) even parity as the vacuum state $\ket{0}$.\footnote{One can prove that $[\Pi,S(r,\theta)\otimes \mathbb{I}_q]=0$ by showing that $[\Pi_c,\chi^\ast a^2 - \chi a^{\dagger 2}]=0$, where $\chi=r e^{i \theta}$ and $\Pi_c \equiv e^{i \pi a^\dagger a}$.
This follows from the identity $e^{-\alpha a^\dagger a} f(a, a^\dagger) e^{\alpha a^\dagger a} = f(a e^\alpha, a^\dagger e^{-\alpha})$ with $\alpha \in \mathbb{C}$ (see, e.g, Problem 1.4 in \cite{Scully_Zubairy_1997}), which implies $\Pi_c^\dagger a^{(\dagger)} \Pi_c = -a^{(\dagger)}$, and so $\Pi_c^\dagger a^{(\dagger)2} \Pi_c = (\Pi_c^\dagger a^{(\dagger)} \Pi_c)^2 = a^{(\dagger)2}$ since $\Pi_c$ is unitary.}
Finally, when expanded in the Fock basis, the target Schr\"{o}dinger-cat-state superposition $\ket{\psi_{\rm target}}$ in Eq. (B4) comprises exclusively even-number basis states of the form $\{\ket{2(2k+1)}\}_{k\in \mathbb{N}_0}$,\footnote{While a standard even cat state $\ket*{C_\alpha^+}$ contains exclusively even Fock states $\ket{2k}$, the periodic properties of the imaginary unit in Eq. (B4) make the superposition state $\ket{\psi_\mathrm{target}}$ retain only those specific components, filtering out the others.} thus requiring the qubit to be in the ground state $\ket{g}$ to satisfy parity conservation.

\subsection{Hybrid strategy}
\label{sec:hybrid}
In our implementation, the GRAPE optimization is initialized using the control field obtained from a prior CRAB optimization \footnote{To obtain a smoother control profile, the optimized GRAPE pulse is subsequently interpolated using a shape-preserving cubic spline (PCHIP interpolation as implemented in \texttt{SciPy}). We verified that the final fidelity exhibits only a weak dependence on both the interpolation scheme and the interpolation time step.}. This hybrid strategy combines the smooth parametrization provided by CRAB with the high-precision local refinement enabled by GRAPE. Since we are interested in targeting a subsystem state (cavity) of a bipartite system (cavity--qubit), the definition of the cost functional slightly
differs in the two methods:
CRAB's cost function relies on the fidelity of the reduced cavity state [Eq. \eqref{eq:fidelity}], while GRAPE, whose gradient computation relies on both the backward-propagated target state and the forward-propagated initial state [Eq. \eqref{eq:grad}], necessarily operates at the level of the full bipartite state.
Despite this difference in the implementation, both approaches target the same physical objective, i.e., the preparation of the desired cavity state, with the qubit acting as an auxiliary degree of freedom.
We emphasize that all fidelities reported in the main text are evaluated with respect to the reduced cavity state [see Eq. {\eqref{eq:fidelity}}], thus ensuring a consistent comparison between the two optimization strategies.

\section{Numerical characterization of the control landscape for Fock-state generation}
To better understand the performance of the CRAB framework and identify the driving regimes where the system most efficiently generates non-classical cavity states, in particular Fock states, we performed a systematic set of numerical tests. Specifically, we evaluated the role of the number of Fourier harmonics, the total evolution time, the frequency sampling range, and the constraint on the total power. We characterized each of these tests in terms of two metrics---final fidelity with respect to the target Fock state $|n=6\rangle$, and the control cost $C_2$ [Eq. (4)]---reporting their mean value and standard deviation over 10 indepedent realizations.

\subsection{Scaling with the number of harmonics} 
\label{sec:scaling_Nf}
We first investigate how the performance of the CRAB optimization depends on the number of Fourier harmonics $N_f$ used in the truncated expansion of the control field in Eq. \eqref{eq:drive_expansion_CRAB}. Increasing $N_f$ enhances the expressivity of the parametrization, but also enlarges the optimization space, potentially hindering convergence.
To quantify this trade-off, we perform a systematic scan over $N_f \in [2, 40]$. For each value of $N_f$, we consider two representative coupling strengths $g/\omega_c = 0.3, 0.6$ and two evolution times $T = 10\,\tau_s, 20\,\tau_s$, where $\tau_s=\pi/2g$ 
sets the characteristic timescale. For each parameter set, we perform 10 independent optimization runs with different random seeds, which affect both the initialization of the Fourier coefficients $\{A_k,B_k\}$ and the sampling of the control frequencies $\{\omega_k\}$ . The frequencies are independently drawn from a uniform distribution $\omega_k/\omega_c \in [0, 6]$, allowing for a statistical characterization of the optimization performance.

The results are shown in Fig.~\ref{fig:fid_amp} for $g/\omega_c=0.3$ and $T=20\,\tau_s$. The average fidelity [Fig.~\ref{fig:fid_amp}(a)] increases with $N_f$ and reaches a plateau around $N_f \simeq 20$, remaining nearly constant thereafter at values close to unity.
For larger values (e.g., $N_f = 40$), a slight degradation is observed, indicating reduced optimization efficiency in higher-dimensional parameter spaces.
At the same time, the variance across realizations decreases with increasing $N_f$, suggesting reduced sensitivity to the specific frequency sampling. Because the frequencies are sampled from a fixed interval, larger values of $N_f$ increase the likelihood of sampling frequencies close to the optimal ones, while frequencies that do not contribute to the optimization can be assigned negligible amplitudes. This results in a more robust optimization landscape and improved reproducibility across realizations.
The average control cost [Fig.~\ref{fig:fid_amp}(e)] increases rapidly with $N_f$ for small $N_f$, reaches a maximum, and then remains approximately constant at lower values for $N_f \gtrsim 10$.
The relatively large fluctuations indicate that different frequency realizations can still lead to substantially different pulse structures.
No qualitative differences are observed upon varying $g$ or $T$ within the explored ranges.
Based on these results, hereafter we set $N_f = 20$, serving as a good compromise between performance (high fidelity and relatively low control cost), robustness (small fidelity fluctuations), and computational cost (a low-dimensional parameter space).

\subsection{Scaling with the total evolution time} 
Next we investigate the dependence of the CRAB optimization performance on the total evolution time $T$.
We perform a systematic scan over $T  \in [5\,\tau_s,\, 40\,\tau_s]$, fixing the coupling strength to $g/\omega_c = 0.3$ and using $N_f = 20$ harmonics, as previously motivated in Sec. \ref{sec:scaling_Nf}. For each $T$, we run 10 independent optimizations with different random seeds, with frequencies sampled uniformly in $[0, 6\,\omega_c]$.

As shown in Fig.~\ref{fig:fid_amp}(b), the achievable fidelity depends on $T$, particularly in the short-time regime. For small $T$, the optimization fails to reach high fidelities, indicating insufficient time to prepare the target state. As $T$ increases, the fidelity rapidly improves and saturates, reaching a plateau beyond which no significant gains are observed.
This behavior indicates the existence of a minimum evolution time required to efficiently exploit the underlying dynamical mechanisms. Beyond this threshold, high fidelities ($F \gtrsim 0.95$) can be consistently achieved.
On the other hand, the average control cost [Fig.~\ref{fig:fid_amp}(f)] exhibits an overall increasing trend with $T$, reflecting a higher cumulative energy cost for longer protocol durations.
Based on these results, hereafter we focus on $T = 20\,\tau_s$, which provides a good compromise between fidelity, control amplitude, and protocol duration.

\subsection{Scaling with the frequency sampling range} 
\label{sec:scaling_bandwidth}
We then investigate the dependence of the optimization performance on the frequency bandwidth used in the CRAB ansatz. The Fourier frequencies are sampled from a uniform distribution $\omega_k \in\big[0, \kappa \, \omega_c \big]$, where $\kappa$ controls the accessible bandwidth.
We scan $\kappa \in [1, 12]$, fixing $N_f = 20$, $T = 20\,\tau_s$, and $g/\omega_c = 0.3$. For each $\kappa$, we perform 10 independent optimizations with different random seeds.

\begin{figure*}[!ht]
    \centering
    \includegraphics[width=\linewidth]{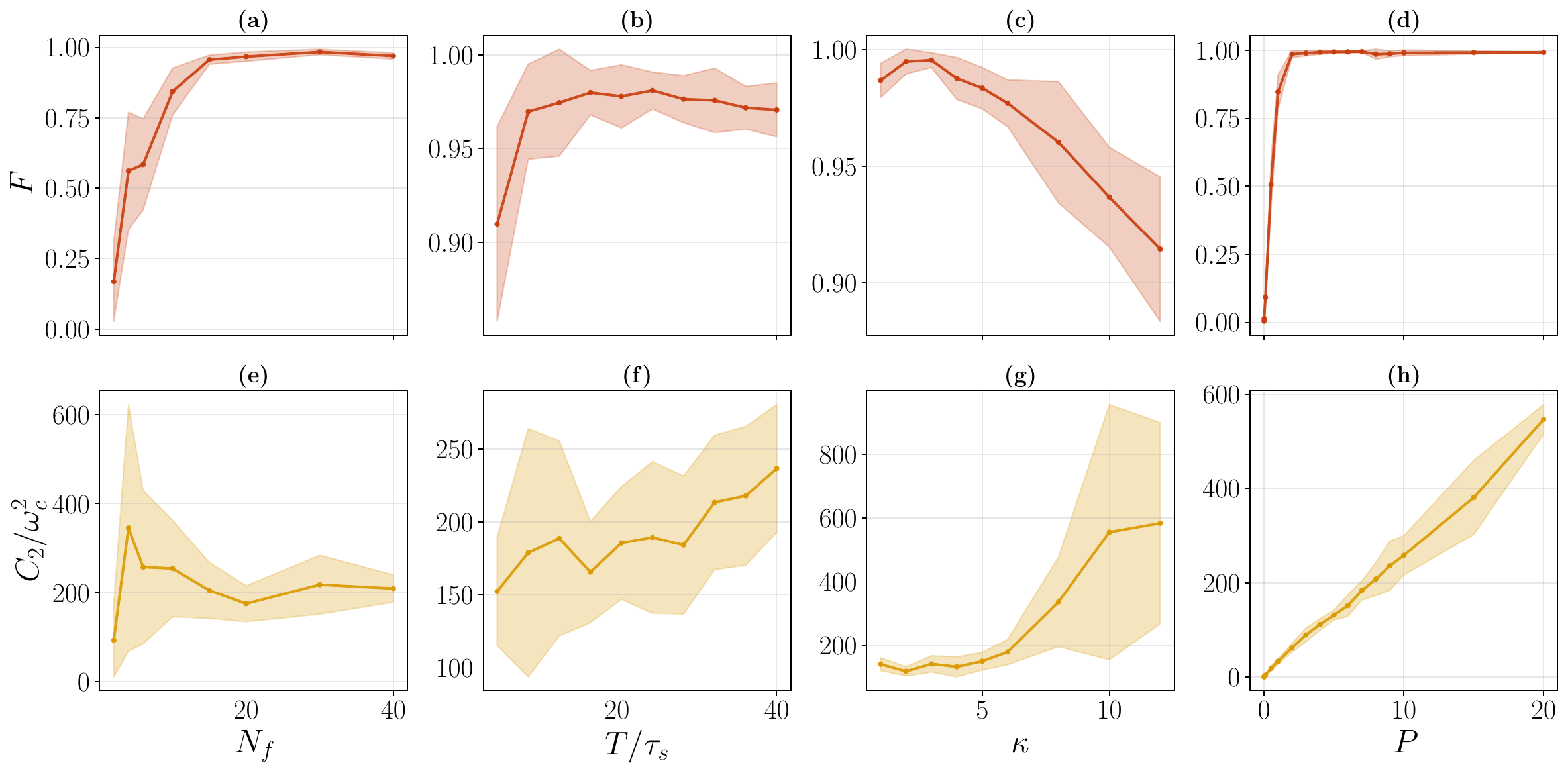}
    \caption{
    Performance of the CRAB optimization as a function of the control parametrization and resource constraints in targeting the Fock state $\ket{n=6}$.
    Panels (a,e): dependence on the number of harmonics $N_f$. Parameters: $g/\omega_c=0.3$, $T=20\,\tau_s$, $\omega_k/\omega_c\in[0, 6]$, $\omega_0/\omega_c=2$. 
    Panels (b,f): dependence on the total evolution time $T$. Parameters: $g/\omega_c=0.3$, $N_f=20$, $\omega_k/\omega_c\in[0, 6]$, $\omega_q/\omega_c=2$.
    Panels (c,g): dependence on the frequency bandwidth parameter $\kappa$, with frequencies sampled in the interval $\omega_k \in [0,\kappa\,\omega_c]$. Parameters: $g/\omega_c=0.3$, $N_f=20$, $T=20\,\tau_s$, $\omega_q/\omega_c=2$.
    Panels (d,h): dependence on the total control power $P$. Parameters: $g/\omega_c=0.3$, $N_f=20$, $T=20\,\tau_s$, $\omega_k/\omega_c\in[0, 3]$, $\omega_q/\omega_c=2$. 
    The top row [(a)--(d)] shows the average final fidelity with respect to the target Fock state $\ket{n=6}$. The bottom row [(e)--(h)] reports the average control cost.
    Error bands denote standard deviations over 10 independent optimization runs with different random seeds.
    }
    \label{fig:fid_amp}
\end{figure*}

As shown in Fig.~\ref{fig:fid_amp}(c), the fidelity strongly depends on the bandwidth. For small $\kappa$ ($\kappa \simeq 1$), the restricted frequency range limits the expressivity of the ansatz, preventing high fidelities. Increasing $\kappa$ improves the performance, with a maximum around $\kappa \simeq 2$--$3$.
However, for larger $\kappa$ the fidelity decreases while fluctuations across realizations increase, reflecting the increased complexity of the optimization landscape: sampling over a wider interval introduces high-frequency components that are not relevant to the target dynamics, enlarging the search space and hindering convergence.
The average control cost [Fig.~{\ref{fig:fid_amp}}(g)] shows an overall increasing trend with $\kappa$. While remaining moderate for $\kappa \lesssim 4$, it grows significantly at larger bandwidths, indicating increasingly demanding control fields with no corresponding improvement in fidelity [Fig.~{\ref{fig:fid_amp}}(c)].
These results indicate that the relevant dynamics are captured within an intermediate spectral window. Accordingly, hereafter we restrict to $\kappa = 2$--$3$, serving as a good compromise between expressivity and optimization efficiency.

\subsection{Constraint on the total control power} 
We now investigate the effect of limiting the overall strength of the control field. Letting $\vb{c} = \{A_k, B_k\}_{k=1,\dots,N_f}$ denote the vector of Fourier coefficients entering the CRAB parametrization [Eq. \eqref{eq:drive_expansion_CRAB}], we define $P = \| \vb{c}\|^2 = \sum_{k=1}^{N_f}(A_k^2+B_k^2)$, which serves as a proxy for the total control power.\footnote{For a control field expressed as a standard Fourier series over a fixed period, Parseval's theorem ensures that $\| \vb{c}\|^2$ is proportional to the time-averaged control power. However, in the CRAB parametrization, due to the truncated and generally randomized-frequency basis, this identification is not exact, and $P$ should instead be interpreted as a heuristic measure of the overall control power.}
In our implementation, we constrain $P$ to limit the overall strength of the control field, thereby reflecting finite control resources. 
At each optimization step, the vector of Fourier coefficients $\vb{c}$ is rescaled so that $\|\vb{c}\|^2 = P$ (see, e.g., \cite{10.1116/5.0190340}) via
\begin{equation}
\vb{c} \rightarrow \sqrt{P}\,\frac{\vb{c}}{\|\vb{c}\|}.
\end{equation}
This constraint is enforced during pulse reconstruction, ensuring that all candidate controls satisfy the same global resource budget.
We scan $P \in [1,20]$, fixing $N_f = 20$, $T = 20\,\tau_s$, and $g/\omega_c = 0.3$.
We consider two bandwidths, $\omega_k/\omega_c \in [0,2]$ and $\omega_k /\omega_c\in [0,3]$,  which, according to the previous analysis in Sec. \ref{sec:scaling_bandwidth}, yield the highest fidelities,
and perform 10 optimizations with different random seeds.

The results are summarized in Fig.~\ref{fig:fid_amp} for $\omega_k/\omega_c\in[0,3]$ (similar results are obtained for $\omega_k/\omega_c\in[0,2]$).
The average fidelity increases with $P$, reaching a plateau close to $F \approx 1$ for $P \gtrsim 2$ [Fig.~\ref{fig:fid_amp}(d)]. Instead, the average control cost increases monotonically [Fig.~\ref{fig:fid_amp}(h)] with $P$, reflecting more demanding control fields at larger $P$.
Overall, these results highlight a trade-off between performance and resource consumption. Small $P$ restricts the accessible control landscape, limiting the achievable fidelity but yielding lower control costs. Large $P$ recovers nearly unconstrained performance at the cost of more demanding controls. An intermediate regime emerges in which high fidelities are obtained with moderate resources, located around $P \simeq 2$--$3$, depending on the bandwidth.

Based on these results, we select as initial guess for subsequent GRAPE optimizations a control pulse with moderate fidelity but low control amplitude (corresponding to $P=2$), providing a favorable compromise between performance and limited control resources.
The GRAPE refinement then enhances the fidelity while allowing for a constrained optimization of the control amplitude within the range of amplitudes obtained in the preceding CRAB optimization.

\section{Details on the dissipation model}
The environment is described as a thermal bath of harmonic oscillators in equilibrium at inverse temperature $\beta$, with Hamiltonian $H_B = \sum_k \omega_k {b}_k^\dagger {b}_k$, where $b_k^{(\dagger)}$ are annihilation (creation) operators of the bath modes.
The system comprises both the cavity and the qubit, with Hamiltonian $H_S(t) = H_R+H_D(t)$ [see Eqs. (1)--(2)].
The system--environment interaction is taken to be linear in the bath coordinates,
\begin{equation}
    H_{I} = V \otimes \sum_k g_k (b_k^\dagger + b_k),
\end{equation}
where $V$ is a system operator and $g_k$ denote the coupling strengths. Following standard results in open quantum systems theory~\cite{breuer2002theory,Manzano,Campaioli,vacchini2024open}, the reduced system dynamics is described by the GKLS master equation~\cite{kossakowski1972nonhamiltonian,lindblad1976generators,gorini1976completely}
\begin{equation}
\frac{d}{dt}\rho_S(t) = -i[H_S(t),\rho_S(t)] 
\nonumber+ 
\sum_\omega \left( L_\omega(t) \rho_S(t) L_\omega^\dagger(t) - \frac{1}{2}\{L_\omega^\dagger(t) L_\omega(t),\rho_S(t)\} \right),
    \label{eq:GKSL_me}
\end{equation}
where $H_S(t)=\sum_\epsilon \epsilon(t) \dyad{\epsilon(t)}$ is the time-dependent system Hamiltonian with instantaneous eigenvalues $\epsilon(t)$ and corresponding instantaneous eigenstates $\ket{\epsilon(t)}$. Accordingly, the Bohr frequencies are defined as $\omega(t) = \epsilon'(t) - \epsilon(t)$, and the Lindblad operators take the form
\begin{equation}
    {L}_\omega(t) = \sqrt{\gamma(\omega)} \sum_{\epsilon'(t) - \epsilon(t) = \omega} V_{\epsilon \epsilon'}(t)\, |\epsilon(t)\rangle \langle \epsilon'(t)|,
\end{equation}
with $V_{\epsilon \epsilon'}(t) = \langle \epsilon(t) | {V} | \epsilon'(t) \rangle$. 

We consider two scenarios in which the environment couples directly either to the cavity mode or to the qubit. In the former case, we take $V = (a + a^\dagger) \otimes \mathbb{I}_q$; in the latter, $V = \mathbb{I}_c\otimes\sigma_x$. Here, $\mathbb{I}_{q(c)}$ denotes the identity operator on the qubit (cavity) Hilbert space. In both cases, however, the Lindblad operators are constructed from the instantaneous energy eigenstates of the joint cavity-qubit system.
Therefore, due to the USC regime of the cavity-qubit interaction, dissipation effectively acts on the whole cavity-qubit system. The transition rates are given by
\begin{equation}
    \gamma(\omega) = \frac{2\pi J(|\omega|)}{1 - e^{-\beta |\omega|}} \left[ \Theta(\omega) + e^{-\beta |\omega|} \Theta(-\omega) \right],
\end{equation}
where $J(\omega)$ is the bath spectral density, $k_B = 1$ and $\Theta(\omega)$ is the Heaviside function.
We assume an Ohmic spectral density with exponential cutoff $J(\omega) = \eta |\omega| e^{-|\omega| / \omega_{B}}$, where $\eta$ sets the system–bath coupling strength and $\omega_B$ is the bath cutoff frequency.

\section{Comments on experimental feasibility}

A promising platform for implementing our proposal is provided by superconducting flux qubits inductively coupled to lumped-element LC resonators through Josephson junctions \cite{yoshihara2017natphys,yoshihara2017pra}. In such systems, resonator and qubit frequencies typically lie in the range
$$
\omega_{c,(q)}/2\pi \sim 1\text{--}10~\mathrm{GHz},
$$
while the large anharmonicity of flux qubits preserves an effective two-level description even under strong driving conditions.

Microwave driving amplitudes exceeding the qubit transition frequency have been experimentally demonstrated using high-speed arbitrary waveform generators with subnanosecond resolution \cite{deng2015prl}. Specifically, driving amplitudes up to $2\pi\times4.78~\mathrm{GHz}$ have been achieved for a qubit transition frequency of $2\pi\times2.288~\mathrm{GHz}$, indicating that the optimized control pulses considered in this work lie within current technological capabilities.

To quantify the speedup enabled by the ultrastrong-coupling regime, we compare our protocol for Fock-states generation with the cavity-state preparation scheme of Ref.~\cite{Hofheinz2008}, based on a superconducting phase qubit coupled to a cavity through the Jaynes--Cummings interaction. For the preparation of the Fock state $\ket{n=6}$ at a cavity frequency $\omega_c/2\pi = 6.57~\mathrm{GHz}$, the protocol of Ref.~\cite{Hofheinz2008} requires approximately
$$
T \simeq 190~\mathrm{ns},
$$
whereas our approach reaches the same target state in only
$$
T = 20\,\tau_s \simeq 2.54~\mathrm{ns}.
$$
This corresponds to a speedup of nearly two orders of magnitude and places the preparation time well below typical superconducting quantum-gate timescales \cite{kjaergaard2019annrev,Krantz2019appphysrev}.
With respect to the generation of Schr\"{odinge}-cat-state superpositions, the scheme reported in Ref.~\cite{PhysRevLett.111.120501} requires a sequence of approximately 13 gates, corresponding to a total duration on the order of microseconds (depending on the experimental implementation). In contrast, our protocol operates on a timescale of approximately $2\,\mathrm{ns}$ in the considered parameter regime (setting $\omega_c/2\pi = 5$ GHz).

Finally, the open-system simulations presented in Figs.~2(b),(e) and 4(b) employ environmental parameters representative of current superconducting-circuit experiments \cite{Niemczyk2010,Fink2008}. In particular, $\omega_B/\omega_c=10$ and $\beta\omega_c=5$ correspond to a temperature of approximately $48~\mathrm{mK}$ for a cavity frequency $\omega_c/2\pi=5~\mathrm{GHz}$.

\FloatBarrier
\bibliography{references}